\documentclass[showpacs,showkeys,aps,twocolumn]{revtex4}
\usepackage{graphicx,amsmath,amsfonts,amssymb}
\usepackage[normalem]{ulem}
\usepackage{color}
\usepackage{times}

\newcommand{\be}{\begin{equation}}
\newcommand{\ee}{\end{equation}}

\newcommand{\bea}{\begin{eqnarray}}
\newcommand{\eea}{\end{eqnarray}}

\begin{document}
\title{Collective Coordinates Theory for Discrete Soliton Ratchets in the sine-Gordon Model}
\author{Bernardo S\'anchez-Rey}
\email{bernardo@us.es} \affiliation{Departamento de F\'\i sica
Aplicada I, E.P.S., Universidad de Sevilla, Virgen de \'Africa 7,
41011, Sevilla, Spain}

\author{Niurka R.\ Quintero}
\email{niurka@us.es} \affiliation{Instituto de Matem\'aticas de la
Universidad de Sevilla (IMUS)} \affiliation{Departamento de F\'\i
sica Aplicada I, E.P.S., Universidad de Sevilla, Virgen de \'Africa
7, 41011, Sevilla, Spain}

\author{J. Cuevas-Maraver}
\email{jcuevas@us.es} \affiliation{Instituto de Matem\'aticas de la
Universidad de Sevilla (IMUS)} \affiliation{Departamento de F\'\i
sica Aplicada I, E.P.S., Universidad de Sevilla, Virgen de \'Africa
7, 41011, Sevilla, Spain}

\author{Miguel A. Alejo}
\email{malejo@impa.br} \affiliation{Instituto Nacional de
Matem\'atica Pura e Aplicada (IMPA), Estrada Dona Castorina 110,
22460-320, Rio de Janeiro, Brazil}

\date{\today}

\begin{abstract}
 A collective coordinate theory is develop for soliton ratchets in the damped discrete
sine-Gordon model driven by a biharmonic force. An {\it ansatz} with
two collective coordinates, namely the center and the width of the
soliton, is assumed as an approximated solution of the discrete
non-linear equation. The evolution of these two collective
coordinates, obtained by means of the  Generalized Travelling Wave
Method, explains the mechanism underlying the soliton ratchet and
captures qualitatively all the main features of this phenomenon. The
theory accounts for the existence of a non-zero depinning threshold,
the non-sinusoidal behaviour of the average velocity as a function
of the difference phase between the harmonics of the driver,  the
non-monotonic dependence of the average velocity on the damping and
the existence of non-transporting regimes beyond the depinning
threshold. In particular it provides a good description of  the
intriguing and complex pattern of subspaces corresponding to
different dynamical regimes in parameter space.
\end{abstract}

\maketitle

\section{Introduction}
Soliton Ratchet~\cite{Marchesoni,Niurka1,Costantini} is a
generalization of the ratchet effect~\cite{Ratchet} to spatially
extended systems described by nonlinear partial differential
equations (PDE). In these systems, nonlinear coherent excitations
play a key role for transport properties and the ratchet effect
appears as a directed motion of solitons under the influence of zero
mean forces, due to the breaking of the spatiotemporal or field
symmetries of the system~\cite{Niurka1,Salerno1,Flach}.

This phenomenon has been observed experimentally in long Josephson
junctions (JJ) devices.  In this system, a spatially asymmetric
sawtooth ratchet potential can be emulated using a inhomogeneous
magnetic field~\cite{Carapella}, giving rise to  drift dynamics of
fluxons (magnetic flux quanta) similar to that observed for
particle-point  ratchets. The break of the spatial symmetry can also
be achieved by properly injecting  an external current~\cite{Beck}
or by introducing a modulation of the coupling between
junctions~\cite{Trias,Shalom}. Another means to obtain a fluxon
ratchet is that of breaking the temporal symmetry. That was
performed in an annular JJ using a bi-harmonic force accomplished
with microwaves~\cite{Ustinov04}.

The mechanism underlying soliton ratchets in continuous systems has
been clarified in detail by a  collective coordinates (CC) theory
\cite{Salerno1,Niurka2,Niurka3,moralesmolina2004,moralesmolina2005}.
In this theory, the center of mass, $X(t)$, and the width, $l(t)$,
of the soliton are  independent dynamical variables and the initial
PDE is reduced to a pair of coupled nonlinear ordinary differential
equations (ODE) for the two CCs. Net soliton motion becomes possible
when the indirect driving resonates with one of the available
frequencies of the soliton internal vibrations. If this resonant
condition is fulfilled, the energy pumped by the driver into the
soliton internal vibration  is converted into net motion through its
coupling with the translational degree of freedom.

Soliton ratchets have also been investigated in discrete
systems~\cite{Salerno2,Cuevas,Gorbach,Chacon,yang}. In contrast with
the continuous case, the interplay between discreteness and
nonlinearity introduces new features such as: non-zero depinning
threshold~\cite{Zolotaryuk}, staircase dependence of the mean
velocity on system parameters, chaotic or  intermittent ratchet-like
dynamics and the intriguing phenomenon of the existence of
non-transporting regimes beyond the depinning
threshold~\cite{Cuevas}. However, most studies are numerical due to
the lack of an adequate analytical approximation for strongly
discrete systems. The aim of this paper is precisely to address this
challenge by developing a two-CC theory that captures all the rich
phenomenology of the discrete soliton ratchets in the
Frenkel-Kontorova model.

The paper is organized as follows. In Sec. II the damped
Frenkel-Kontorova model driven by external bi-harmonic force is
introduced. An \textit{ansatz} with two collective coordinates is
suggested as an approximated solution of this system, and the ODEs,
which govern the evolution of the CCs, are obtained by means of the
Generalized Travelling Wave Method (GTWM). Section III is devoted to
simulations of the original PDE and their comparison with the
numerical simulations of the ODEs for the two CCs. The dependence of
the soliton mean velocity on various system parameters is studied in
detail and particular emphasis is laid on the transition from the
very discrete limit to the continuous limit. Section IV summarizes
the main conclusions of the paper.

\section{Model and collective coordinates theory}
   We will focus on the paradigmatic case of  the
damped discrete sine-Gordon (sG) system
\cite{Braun1998,Kivshar,sGbook}
\begin{equation}
\ddot{\phi}_n-\kappa \Delta_d \phi_n +\frac{dV(\phi_n)}{d \phi_n}=
-\alpha \dot{\phi} _n+F(t),\quad n=1,2,...,N  .\label{eq:sG}
\end{equation}
which can be used, for instance, to describe a parallel array of JJs
\cite{Falo} or to model a circular array of underdamped JJs
 when $F$ is a constant force \cite{usti1993}. Here, $\phi_n$ is a scalar field,
$\dot{\phi}$ is the derivative with respect to time, $\Delta_d
\phi_n\equiv \phi_{n+1}+\phi_{n-1}-2\phi_n$ is the discrete
Laplacian, $\kappa$ is the coupling constant that measures the
discreteness of the lattice, $\alpha$ is the dissipation parameter,
$F(t)$ is a time-periodic external driving, and $\frac{dV(\phi_n)}{d
\phi_n}=\sin(\phi_n)+\lambda \cos(2 \phi_n)$ is the derivative of a
spatially-periodic potential, which will be spatially asymmetric in
case parameter $\lambda\ne 0$. For $F=0$ and $\lambda \ne 0$, a
mechanical analogue  of Eq. (\ref{eq:sG}) in terms of  a chain of
double pendulum was given in \cite{ms85}.  All these magnitudes and
parameters are in dimensionless form.

 In this system, net motion of topological nonlinear
excitations (kinks or antikinks) can arise when the symmetries of
Eq. (\ref{eq:sG}), which relate kink solutions with opposite
velocities, are broken. In order to describe the resulting ratchet
dynamics, a CC theory is presented  based on the idea that
perturbations of the system act essentially on the center of mass,
$X(t)$, and the width, $l(t)$, of the unperturbed discrete kink. In
the same spirit as that of the Rice \textit{ansatz} \cite{Rice} for
the continuous sG equation, the discrete kink is approximated with
the \textit{ansatz} \cite{cisneros2008}
\begin{equation}
\phi_n(t)=4 \arctan \big[\exp(\theta_n)\big] \, , \label{eq:ansatz}
\end{equation}
where
\begin{equation}\theta_n=\frac{\kappa^{-1/2} \; n-X(t)}{l(t)}
\end{equation}
and  equations of motion for the CCs are derived using the so-called
GTWM. A more complex \textit{ansatz}, which includes a modulation of
the Goldstone mode was used in \cite{cisneros2008} to study the
evolution of a propagating kink in the Frenkel-Kontorova model
without perturbations. Since only the case of one kink is to be
considered, periodic boundary conditions $\phi_{n+N}(t)=
\phi_n(t)+2\pi$ are also adopted, which correspond, for instance, to
a circular array of JJs with only one fluxon
\cite{Ustinov04,Ustinov92}.

The GTWM was introduced in a general way in Ref. \cite{Mertens} and
was  successfully applied to study the zero temperature dynamics
\cite{Mertens} and thermal diffusion \cite{Kamppeter} of magnetic
vortices in the two-dimensional anisotropic Heisenberg model. Closer
to our problem, the GTWM has also been used to explain various
phenomena of kink dynamics in continuous $\varphi^4$ and sG models
\cite{Niurka4}. In short, according to the GTWM, given an
\textit{ansatz} with $M$ collective coordinates ${A_1,A_2,...,A_M}$,
the procedure to obtain the evolution equation for each CC, $A_j$
($j=1,\cdots, M$), involves three steps: $i)$ insert the
\textit{ansatz} in the  equation that governs the dynamic of the
field $\phi_n$; $ii)$ multiply the resulting equation by
$\frac{\partial \phi_n}{\partial A_j}$; and  $iii)$ integrate over
the spatial coordinate (sum over all elements $n$ of the lattice in
the discrete case). Thus, applying this prescription, after
straightforward but cumbersome algebra, the following equations for
$X(t)$ and $l(t)$ are attained:

\begin{widetext}
\begin{gather}
I_4 \frac{\ddot X}{l}+I_5 \frac{\ddot l}{l} - (I_3+2 I_5) \frac{\dot
l^2}{l^2}-I_1 \left(\frac{\dot X^2}{l^2}+1 \right)-2(I_2+I_4)
\frac{\dot X \dot l}{l^2} = - I_5 \alpha \frac{\dot l}{l} -I_4
\alpha \frac{\dot X}{l}-I_6 F(t)+ \frac{\lambda}{2} I_7
+\left(I_1+\frac{I_8}{12 \kappa l^2} \right) \frac{1}{l^2},
\label{eq:X} \\
I_5 \frac{\ddot X}{l}+I_{10} \frac{\ddot l}{l} - (I_{9}+2 I_{10})
\frac{\dot l^2}{l^2}-I_2 \left(\frac{\dot X^2}{l^2}+1
\right)-2(I_3+I_5) \frac{\dot X \dot l}{l^2} = - I_{10} \alpha
\frac{\dot l}{l} -I_5 \alpha \frac{\dot X}{l}-I_{11} F(t) +
\frac{\lambda}{2} I_{12} +\left(I_2+\frac{I_{13}}{12 \kappa
l^2}\right) \frac{1}{l^2}, \label{eq:l}
\end{gather}
\end{widetext}
where the terms $I_j\; (j=1,...,13)$  are functions of $X(t)$ and
$l(t)$, that can be defined by the following sums
\begin{equation}
I_1=  \sum_{n=1}^{N} \frac{d \phi_n}{d\theta_n} \sin \phi_n=
-4 \sum_{n=1}^{N} \frac{\sinh \theta_n}{\cosh^3 \theta_n},\label{eq:I1}
\ee \be
 I_2= \sum_{n=1}^{N} \frac{d \phi_n}{d\theta_n} \theta_n
\sin \phi_n
=-4 \sum_{n=1}^{N} \theta_n \frac{\sinh \theta_n}{\cosh^3 \theta_n},\label{eq:I2}
\ee \be
 I_3= \sum_{n=1}^{N} \frac{d \phi_n}{d\theta_n} \frac{d^2
\phi_n}{d\theta^2_n} \theta_n^2=-4 \sum_{n=1}^{N} \theta_n^2 \frac{\sinh \theta_n}{\cosh^3 \theta_n},\label{eq:I3}
\ee \be
I_4= \sum_{n=1}^{N} \Big(\frac{d \phi_n}{d\theta_n}\Big)^2=4 \sum_{n=1}^{N} \frac{1}{\cosh^2 \theta_n},\label{eq:I4}
\ee \be
I_5= \sum_{n=1}^{N} \Big(\frac{d \phi_n}{d\theta_n}\Big)^2 \theta_n=4 \sum_{n=1}^{N} \frac{\theta_n}{\cosh^2 \theta_n},\label{eq:I5} 
\ee \be
I_6= \sum_{n=1}^{N} \frac{d \phi_n}{d\theta_n}=2
\sum_{n=1}^{N} \frac{1}{\cosh \theta_n},\label{eq:I6} \ee \be I_7=
\sum_{n=1}^{N} \frac{d \phi_n}{d\theta_n} \cos 2\phi_n=I_6-16
\sum_{n=1}^{N} \frac{\sinh^2 \theta_n}{\cosh^5
\theta_n},\label{eq:I7}
\ee \be %
 I_8= \sum_{n=1}^{N} \frac{d \phi_n}{d\theta_n} \frac{d^4 \phi_n}{d\theta^4_n}=I_1-24
\sum_{n=1}^{N} \frac{\sinh \theta_n}{\cosh^5 \theta_n},\label{eq:I8}
\ee  \be %
I_{9}= \sum_{n=1}^{N} \frac{d \phi_n}{d\theta_n} \frac{d^2
\phi_n}{d\theta^2_n} \theta^3_n=-4 \sum_{n=1}^{N}
\theta_n^3\frac{\sinh \theta_n}{\cosh^3 \theta_n},\label{eq:I9}
\ee \be %
I_{10}= \sum_{n=1}^{N} \Big(\frac{d \phi_n}{d\theta_n}\Big)^2
\theta^2_n=4 \sum_{n=1}^{N} \frac{\theta_n^2}{\cosh^2
\theta_n},\label{eq:I10}
\ee \be %
I_{11}= \sum_{n=1}^{N} \frac{d \phi_n}{d\theta_n}
\theta_n=2\sum_{n=1}^{N} \frac{\theta_n}{\cosh
\theta_n},\label{eq:I11}
\ee \be %
I_{12}= \sum_{n=1}^{N} \frac{d \phi_n}{d\theta_n} \theta_n \cos 2\phi_n=I_{11}-16 \sum_{n=1}^{N} \theta_n \frac{\sinh^2 \theta_n}{\cosh^5 \theta_n},\label{eq:I12}\\
\ee \be I_{13}= \sum_{n=1}^{N} \frac{d \phi_n}{d\theta_n} \frac{d^4
\phi_n}{d\theta^4_n} \theta_n=I_2-24 \sum_{n=1}^{N} \theta_n
\frac{\sinh \theta_n}{\cosh^5 \theta_n}. \label{eq:I13}
\ee

All these sums are periodic functions of $X(t)$ and, therefore,  can
be expressed as Fourier series (see Appendix A). Numerically, one
can check the fast convergence of the series, being the whole series
very accurately approximated by the first harmonic. Moreover, one
finds that $I_5$, $I_{11}$ and $I_{12}$ are much smaller than the
other $I_j$ terms (see Fig. \ref{fig7} in the appendix).  By bearing
in mind these two additional approximations, the following equations
of motion for the CCs are finally obtained:
\begin{widetext}
\begin{gather}
    \ddot X+\bigg[ J_1(\kappa,l) \frac{\dot{l}^2}{l}- J_2(\kappa,l) \dot{X}^2- J_3(\kappa,l)\bigg]
\sin(2 \pi\sqrt{\kappa}\,  X)- \frac{\lambda}{2} J_4(\kappa,l)
\cos(2\pi\sqrt{\kappa}\,
X)-\frac{\dot X\dot l}{l}= - \alpha \dot X- \frac{\pi}{4} l F(t) \; , \label{eq:Xeq}\\
\ddot l-\frac{\dot l^2}{2 l}+ \frac{6}{\pi^2}\frac{\dot
X^2}{l}+J_5(\kappa,l) \dot X\dot{l} \sin(2 \pi\sqrt{\kappa}\, X) =
-\alpha \dot l+J_6(\kappa,l) , \label{eq:leq}
\end{gather}
\end{widetext}
where
\begin{widetext}
\begin{eqnarray}
J_1(\kappa,l)&=& \frac{\pi} {4 \sinh^3(\pi^2 \sqrt{\kappa}\, l)}
\Big[ 2+3 \pi^4 \kappa l^2 + (-2+\pi^4\kappa\, l^2)\cosh (2\pi^2
\sqrt{\kappa}\, l)\Big] ,\\
J_2(\kappa,l)&=& \frac{2 \pi^3 \kappa l}{\sinh(\pi^2 \sqrt{\kappa}\, l)} , \\
J_3(\kappa,l)&=&  J_2(\kappa,l) l^2+\frac{\pi^3(3+12\kappa l^2+2 \pi^2\kappa l^2)}{6 l \sinh(\pi^2 \sqrt{\kappa}\, l)} \\
J_4(\kappa,l)&=&\frac{4 \pi^3 \kappa l^3(2 \pi^2 \kappa l^2-1)}{3
\cosh(\pi^2 \sqrt{\kappa}\, l)}, \\
J_5(\kappa,l)&=& \frac{6 \pi \sqrt{\kappa}} {\sinh^3(\pi^2
\sqrt{\kappa}\, l)} \Big[ 3 \pi^2 \sqrt{\kappa}\, l + \pi^2
\sqrt{\kappa}\, l \cosh (2\pi^2 \sqrt{\kappa}\, l)
-2 \sinh(2\pi^2 \sqrt{\kappa}\, l)\Big] \\
J_6(\kappa,l)&=&\frac{3[1+4\kappa l^2(1-l^2)]}{2\pi^2\kappa l^3}.
\end{eqnarray}
\end{widetext}

Notice that the validity of the dynamical Eqs. (\ref{eq:Xeq}) and
(\ref{eq:leq}) for the CCs is not restricted to the case of
sinusoidal forces. They are also valid for constant forces or for
any generic time-dependent force. Futhermore, in the equation for
the kink center of mass (\ref{eq:Xeq}), one can identify the
mechanism responsible for the ratchet phenomenon. Indeed, in the
right-handside of that equation, the external force, $F(t)$, appears
coupled to the  internal degree of freedom of the kink, $l(t)$,
giving rise to an effective force $l(t)F(t)$. When $l(t)$ contains
at least one of the harmonics of $F(t)$ and the difference of phases
between these functions satisfy certain conditions, then the average
effective force $\langle l(t)F(t) \rangle \ne 0$ and net kink motion
appears.

By taking  the limit $\kappa\rightarrow \infty$ in Eqs.
(\ref{eq:Xeq}) and (\ref{eq:leq}), one recovers the CC equations in
the continuous limit \cite{Niurka2}
\begin{eqnarray}
\ddot X-\frac{\dot X\dot l}{l}&=& - \alpha \dot X- \frac{\pi}{4} l F(t) \; , \label{eq:cX} \\
2 l\ddot l-\dot l^2+ \frac{12}{\pi^2} \dot X^2 &=& -2\alpha l \dot
l+\frac{12}{\pi^2} (1-l^2)  \label{eq:cl}.
\end{eqnarray}
For a given external force $F(t)$, this two-ODE system can be
studied using  perturbative expansion \cite{Niurka2}. In particular,
for a bi-harmonic force of the form
\begin{equation}
F(t)=\epsilon [\cos(\omega t) +\cos(2\omega t+\varphi)],
\label{eq:biF}
\end{equation}
at first order  in $\epsilon$ and after a transient time, the
average velocity is given by
\begin{equation}
\langle v\rangle  \equiv \lim_{t \to \infty} \frac{1}{t}
\int_{0}^{t}\, \dot{X}(\tau) d\tau \nonumber 
\ee \be \simeq \frac{3\pi\epsilon^3}{D} \Bigg[ \frac{2
\sin(\varphi+\chi+\delta_1)}{\sqrt{\alpha^2\omega^2+(\Omega_R^2-\omega^2)^2}}-
\frac{\sin(\varphi+\chi-\delta_2)}{\sqrt{4
\alpha^2\omega^2+(\Omega_R^2-4 \omega^2)^2}}\Bigg] , \label{eq:v}
\ee
where $D=256 (\alpha^2+\omega^2) \sqrt{\alpha^2+4 \omega^2}, \,
\Omega_R^2= 12/\pi^2$ is the so-called Rice's frequency, $\chi=2
\arctan(\omega/\alpha)-\arctan(2 \omega/\alpha)$, and
$\delta_m=\arctan\left(\frac{\Omega_R^2-m^2\omega^2}{m\alpha\omega}\right)$.
This expression correctly explains  why, in the continuous limit and
for small amplitudes $\epsilon$, $\langle v\rangle$ is proportional
to $\epsilon^3$, depends sinusoidally on the difference phase
$\varphi$ \cite{quintero2013}, decreases with the frequency
$\omega$, and can exhibit current reversals by  properly varying the
damping or the frequency.

\section{Numerical study}
The aim of this section is to check the validity of the CC theory by
numerically integrating  the system of coupled ODEs
(\ref{eq:Xeq})-(\ref{eq:leq}) and by comparing the results with
direct simulations of the discrete sG lattice equation
(\ref{eq:sG}). We will focus on the case of ratchets of discrete
kinks driven by the bi-harmonic force (\ref{eq:biF}) of period
$T=2\pi/\omega$.  As it is well known, in the damped sine-Gordon
equation, this bi-harmonic force breaks the temporal symmetry
$F(t)=-F(t+T/2)$ and can be used to induce net kink transport.
Consequently,  the asymmetry parameter $\lambda$ will heceforth be
taken equal to zero hereafter.

In order to carry out the comparison between the two approaches, we
need to determine the center of mass and the width of the sG kink
profile, $\phi_n$, which map the collective coordinates $X$ and $l$,
respectively. Those variables are defined as:
\begin{eqnarray}
X &=&\Delta\frac{\sum_n n(\phi_{n+1}-\phi_{n-1})^2}{\sum_n(\phi_{n+1}-\phi_{n-1})^2},\\
l &=& \Delta\frac{\sqrt{12}}{3}\left[\frac{\sum_n
n^2(\phi_{n+1}-\phi_{n-1})^2} {\sum_n(\phi_{n+1}-\phi_{n-1})^2}-X^2
\right]^{1/2}
\end{eqnarray}
where $\Delta=1/\sqrt{\kappa}$ is the discretization parameter.

A first and important test for the CC theory is the estimation of
the depinning threshold, i.e., the curve on the $(\kappa, \epsilon)$
plane that separates pinned states from the transporting zone in
which moving kinks exist \cite{Zolotaryuk}. In Fig. \ref{fig1} a
blue continuous line represents the exact depinning curve obtained
from the sG equation (\ref{eq:sG}) for a frequency $\omega=0.1$. It
is computed  by starting from a pinned state and increasing
$\epsilon$, for a fixed value of $\kappa$, until the critical value
for which the kink starts to propagate is found. The same numerical
experiment with the ODEs for the CCs leads to the red circles, which
fit  the exact depinning curve very well. As expected, the critical
driving amplitude for depinning tends to zero when
$\kappa\rightarrow\infty$ and the discreteness effects disappear,
while in the anticontinuum limit $\kappa\rightarrow 0$, the
depinning barrier becomes stronger and larger values of $\epsilon$
are necessary to overcome this barrier.

\begin{figure}
\includegraphics[width=6.0cm]{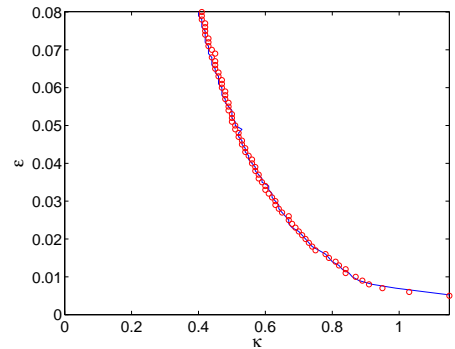}
\caption{(Color online) Depinning curve in the plane $(\kappa,
\epsilon)$  for $\omega=0.1$, $\alpha=0.1$ and  $\varphi=0$. Below
the curve no mobile kinks exist. The area above the curve
corresponds to moving kinks with nonzero average velocity. The
depinning threshold computed from simulation of the discrete sG Eq.\
(\ref{eq:sG}) (blue solid line) is very well fitted by the results
obtained from numerical solutions of the CC equations (red
circles).}\label{fig1}
\end{figure}

    Another distinctive feature of kink ratchets in discrete systems
that is captured by the CC approximation is the piece-wise
dependence of the mean velocity on system parameters. For instance,
in contrast with the continuous case in which  $\langle v\rangle$
has a sinusoidal dependence on the phase difference $\varphi$ for
small amplitudes $\epsilon$, in the discrete case $\langle v\rangle$
becomes zero in intervals whose length increases as the damping is
raised \cite{Salerno2}. Figure \ref{fig2}(a) shows the dependence of
the mean kink velocity on $\varphi$ for $\kappa=2$ and $\alpha=0.1$.
Since this lies not far from the continuous limit, the function
$\langle v\rangle (\varphi)$ computed from the sG equation
(\ref{eq:sG}) (blue continuous line) clearly resembles  a sinusoidal
function. The results obtained with the CC theory, plotted with red
circles, provide a good approximation and capture  the size of the
interval without net transport very well.

    When $\kappa$ is decreased, $\langle v\rangle (\varphi)$ loses its
resemblance with the sine function and the dependence on $\varphi$
takes a more complicated shape, as shown in Fig. \ref{fig2}(b) for
$\kappa=1$. One can observe the existence of many pinning intervals,
and also the existence of many resonant plateaux at $\langle
v\rangle= \pm \Delta/T$, which correspond to kink motion that is
perfectly locked to the external driver frequency such that the kink
moves exactly one site per driving period. These features could be
related with the loss of the regularity of the functional ratchet
velocity \cite{cuesta2013}. The prediction of the CC theory is
quantitatively poorer than in Fig. \ref{fig2}(a). This is not
surprising because the \textit{ansatz} used is based on the exact
solution of the unperturbed system (\ref{eq:sG}) on the continuous
limit. But interestingly, the CC approximation provides a reasonable
description of what happens when one approaches the very discrete
limit.

\begin{figure}
\begin{center}
\includegraphics[width=6.0cm]{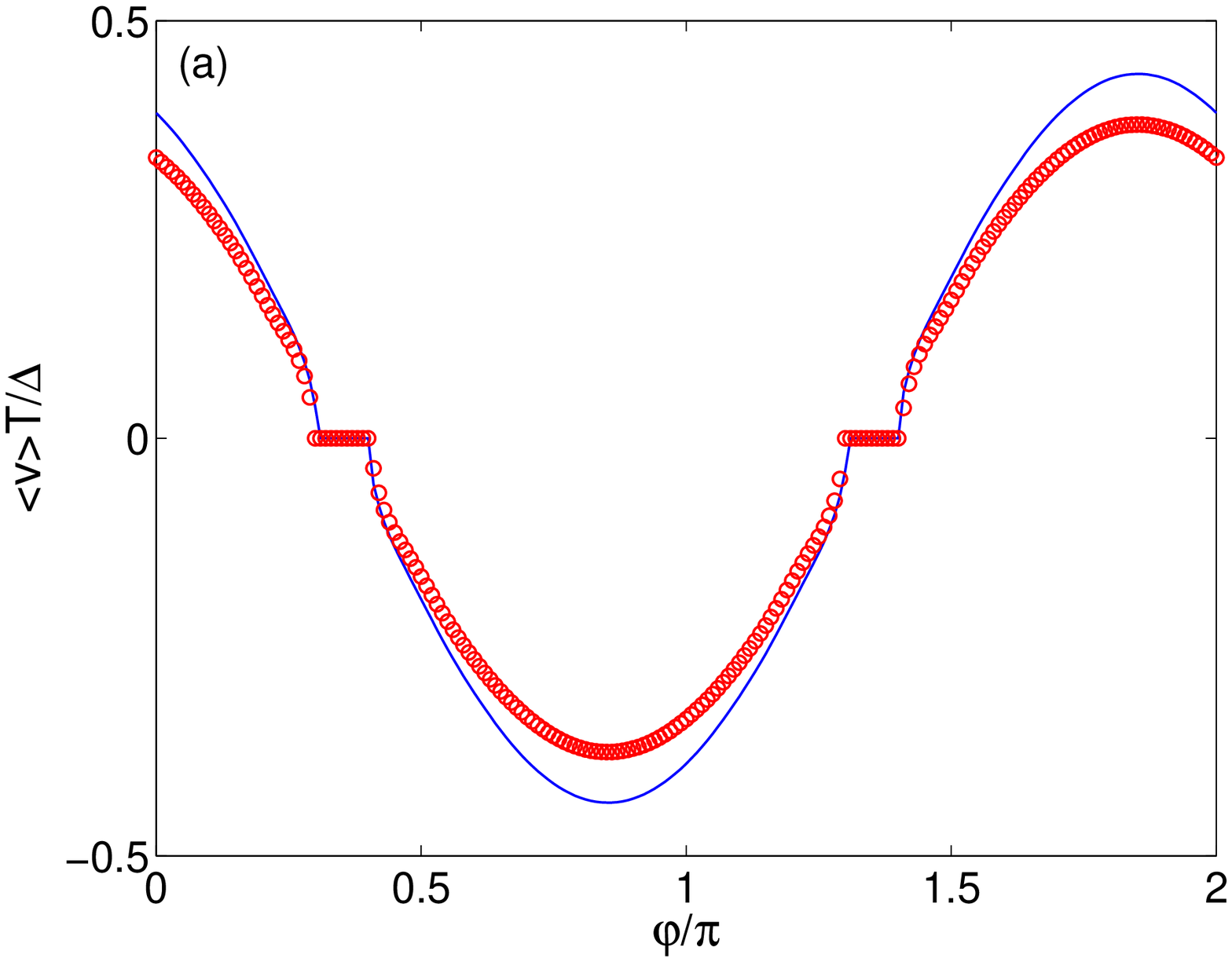}\\
\includegraphics[width=6.0cm]{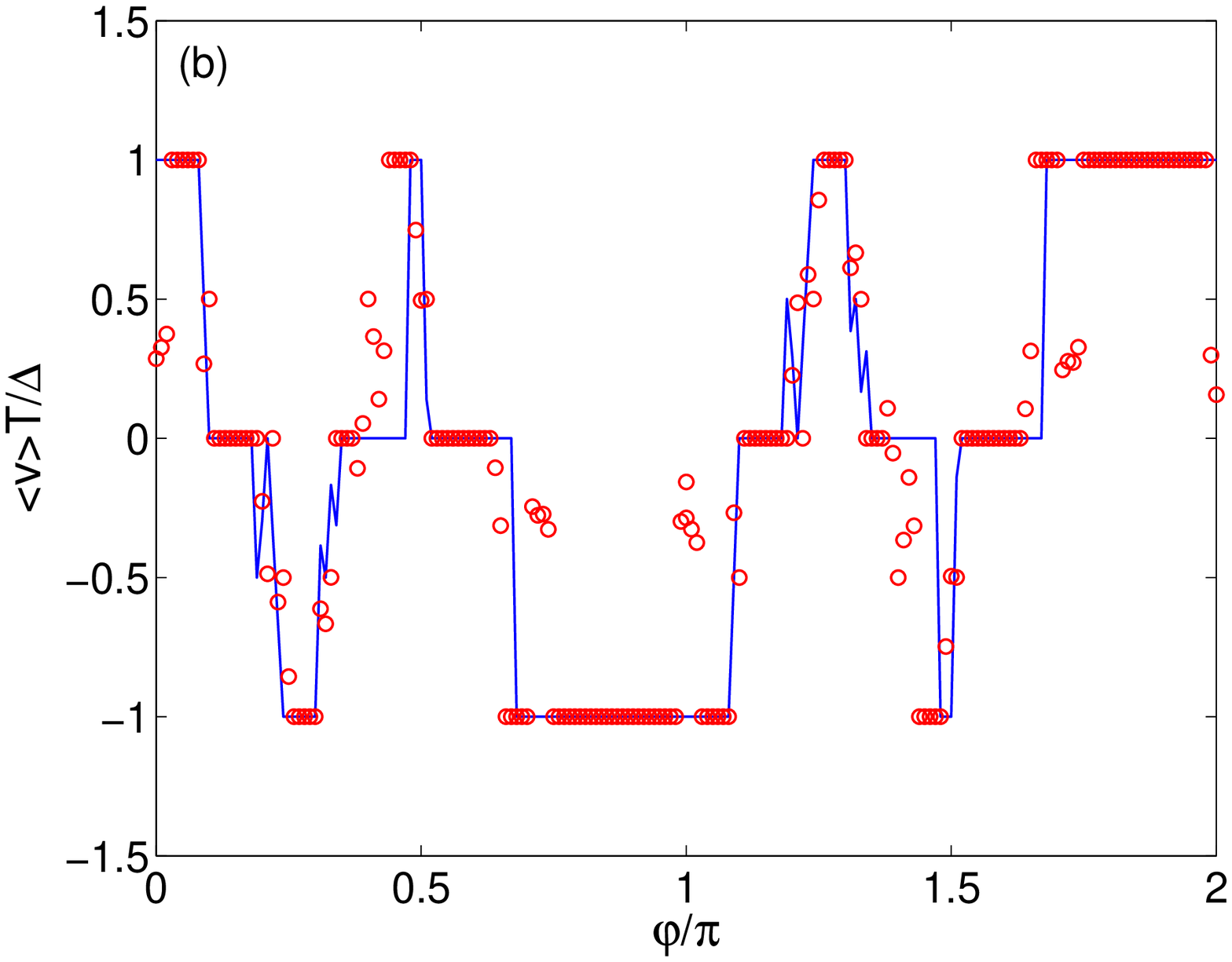} \\
\end{center}
\caption{(Color online) Dependence of the average kink velocity on
the phase difference for $\kappa=2$ (a),  and $\kappa=1$ (b). The
blue continuous line represents results obtained from simulations of
the sG equation (\ref{eq:sG}). Red circles correspond to results
obtained from numerical integration of the CC equations. Other
parameters are $\omega=0.1, \epsilon=0.05, \alpha=0.1$.}
\label{fig2}
\end{figure}

In Fig.\ref{fig3}(a) the transition between the anti-continuous and
the continuous limit in the sG system (\ref{eq:sG}) is studied in
detail, whereas in Fig.\ref{fig3}(b), this transition is considered
through solving the CC equations of motion. Remarkably, in spite of
the very complex scenario described below, the CC theory captures
all the existing dynamical regimes and provides a reasonable
qualitative picture of the whole transition. Along the transition
path different dynamical regimes appear. For very small $\kappa$,
below the depinning threshold, the center of the kink remains pinned
in a potential well oscillating around its minimum (see blue
$\times$). Above the threshold, regular transport corresponding to
periodic (or quasi-periodic) kink trajectories dominates (dots). In
this case, after an integer number $\ell$ of periods $T$, the kink
travels $k$ sites (or very approximately $k$ sites) such that
\begin{equation}
\langle v\rangle= \frac{k \Delta}{\ell T}. \label{eq:resonances}
\end{equation}

Many sudden appearances (and disappearances) of chaotic attractors
(red $+$) leading to diffusive transport are found. The magenta
continuous line represents the analytical estimation (\ref{eq:v}) of
the average velocity at the continuous limit. Note that for $\kappa
\gtrsim 1.5$, we are already very close to the continuous limit.

\begin{figure}
\begin{center}
\includegraphics[width=6.0cm]{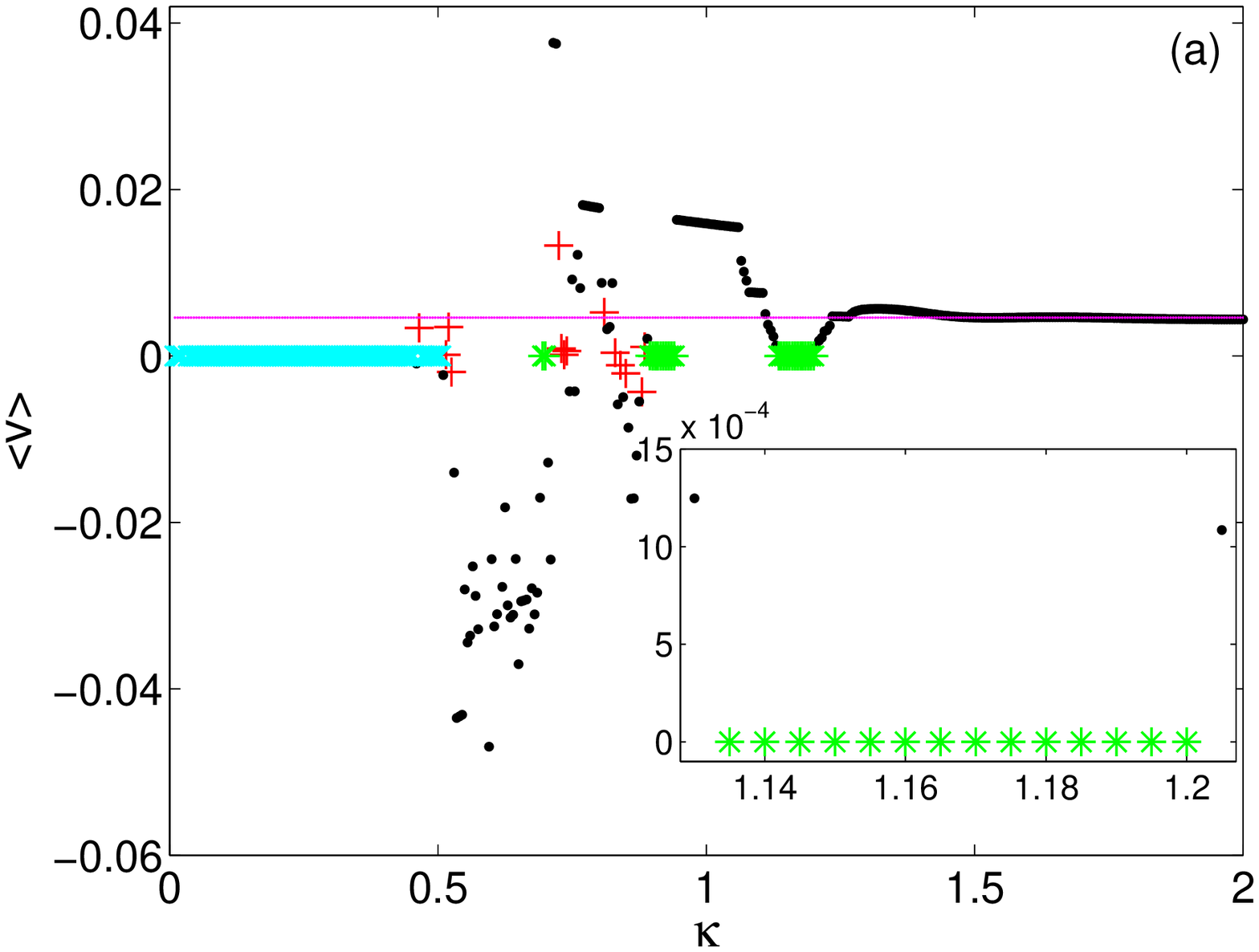}\\
\includegraphics[width=6.0cm]{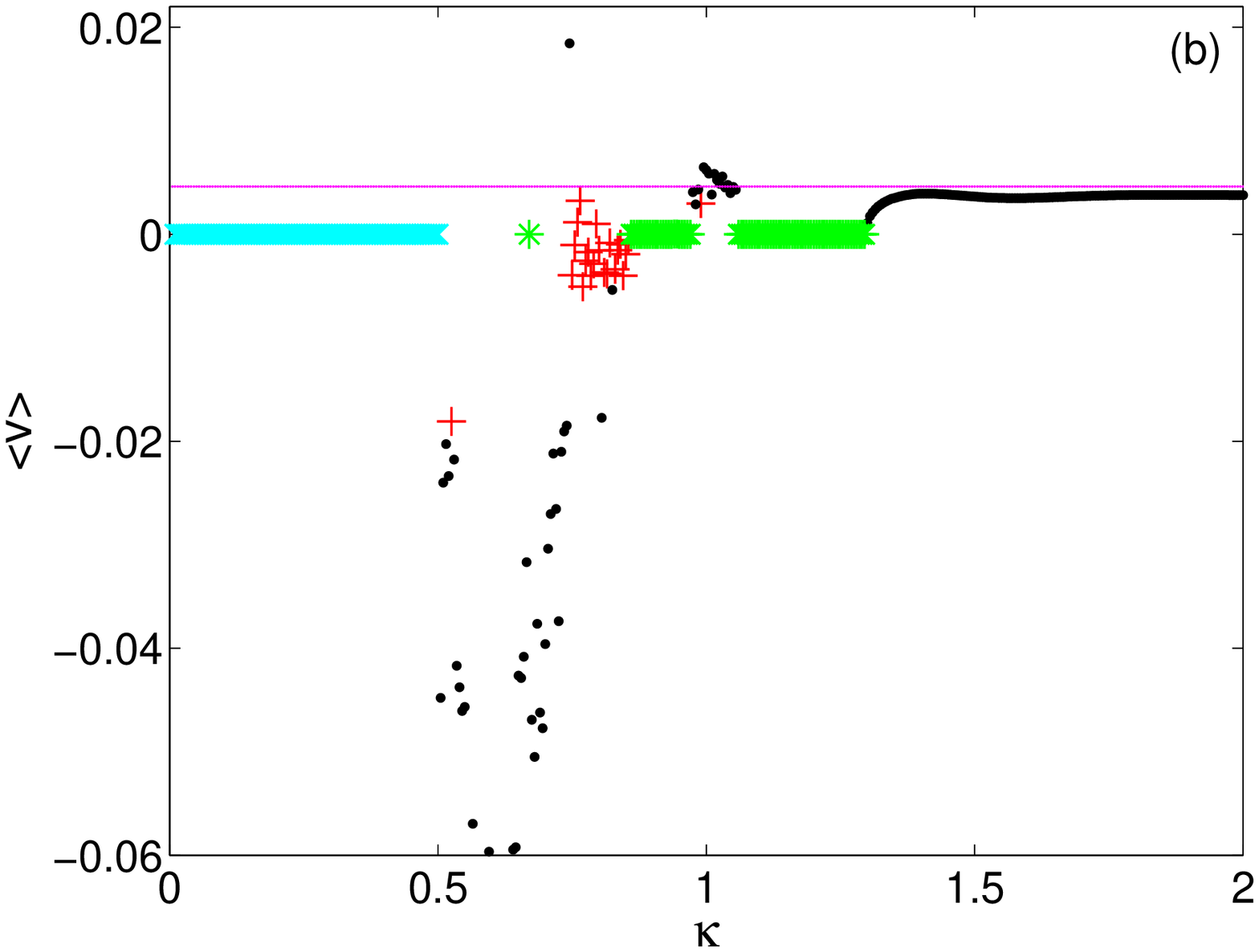} \\
\end{center}
\caption{(Color online) Average kink velocity versus coupling
constant. Panels (a) and (b) correspond to the results from
simulations of the sG system (\ref{eq:sG}) and from the equations of
motion for the CCs (\ref{eq:Xeq})-(\ref{eq:leq}), respectively.
Symbols refer to different dynamical regimes: periodic or
quasiperiodic regular transport (black $\bullet$), chaotic diffusive
transport (red $+$), pinned states (blue $\times$) and rotating
states (green $\ast$). Inset shows more detailed behavior around
$\kappa=1.16$. Other parameters: $\omega=0.1, \epsilon=0.05,
\alpha=0.1$ and $\varphi=0$.} \label{fig3}
\end{figure}

Notice also the existence of significative intervals of
non-transporting {\em rotating} states (green asteriks), which are
far from the depinning threshold. These correspond to periodic
orbits in which the kink center oscillates with an amplitude of
typically several sites of the lattice, i.e., during a period $T$
the kink moves $k$ sites forward and afterwards moves backward,
coming back exactly to its starting position. The dynamic of this
regime is similar to that of a pendulum with {\em oscillating
rotations} (i.e. it rotates $k$ times in one direction, then $k$
times afterwards in the opposite direction, and so forth), contrary
to the pinning states, which resemble a librating pendulum. The
dynamics and Fourier spectra of both regimes are compared in Fig.
\ref{fig4}. Notice that when the kink is pinned (top panels), then
the center of mass oscillates with two harmonics of similar
amplitude and the width is almost constant (note the scale on the Y
axis). For this reason, the effective force acting on the center of
the kink, which is proportional to $l(t) F(t)\approx 1.21 F(t)$  has
practically zero average. On the other hand, in the rotating regime
(bottom panels), the second harmonic of the center of mass has lost
weight, width fluctuations are significative (about $5\%$ of the
average width) and the second harmonic of the width has gained an
importance that was lost by the fourth harmonic. Here, although
$l(t)$ oscillates with at least one of the frequencies of the driver
$F(t)$ (first resonance condition), the phase difference between
these functions is such that $\langle l(t) F(t) \rangle=0$.

Another significant difference arises when analyzing the Floquet
spectrum which enables  the spectral stability and internal modes of
periodic orbits to be determined. Floquet analysis consists firstly
of introducing a perturbation $\xi_n$ to a given solution
$\phi_{n,0}$ of the lattice equations (\ref{eq:sG}). The equation
for the perturbation therefore reads:
\begin{equation}
    \ddot \xi_n-\kappa \Delta\xi_n+(\cos \phi_n)\xi_n+\alpha\dot\xi_n=0.
\end{equation}
The stability properties are given by the spectrum of the Floquet
operator $\mathcal{M}$ (whose matrix representation is called {\em
monodromy}) defined as:
\begin{equation}
    \left(\begin{array}{c} \{\xi_{n}(T)\} \\ \{\dot\xi_{n}(T)\} \\ \end{array}
    \right)=\mathcal{M}\left(\begin{array}{c} \{\xi_{n}(0)\} \\ \{\dot\xi_{n}(0)\} \\ \end{array}
    \right).
\label{cheqn32}
\end{equation}
The $2N$ monodromy eigenvalues $\Lambda=\exp(\mathrm{i}\theta)$ are
dubbed as {\em Floquet multipliers}, and $\theta$ are denoted as
{\em Floquet exponents}. The solution is orbitally stable if all the
multipliers lie on or are inside the unit circle. As shown in
\cite{Marin}, if $\Lambda$ is a multiplier, then so are $\Lambda^*$,
$\rho/\Lambda$ and $\rho/\Lambda^*$, with
$\rho=\exp(-\pi\alpha/\omega)$. Consequently, in the absence of
bifurcations (in our model, for small $\kappa$ \cite{Zolotaryuk}),
all the eigenvalues lie within a circle with radius $\rho$.

As shown in Ref. \cite{Zolotaryuk}, pinned states become mobile by
means of period-doubling or tangent bifurcations, leading the former
to chaotic states and the latter to phase-locked transporting
states. Period-doubling (tangent) bifurcations are caused by Floquet
multipliers departing at $\Lambda=-1$ ($\Lambda=+1$). In Fig.
\ref{fig5}(a), the dependence of the moduli $|\Lambda|$ of the
Floquet multipliers are depicted with respect to $\kappa$ for the
pinning states of Fig. \ref{fig3}(a). It is observed that the kink
remains stable until $\kappa\approx 0.506$. The transition to a
transporting state is mediated by an eigenvalue with exponent
$\theta=0$ that abruptly departs from the circle of radius $\rho$;
that is, the Floquet multipliers stay the whole interval on (or very
close to) the $\rho$-circle. However, in the rotating states, there
is always a multiplier with $\theta=0$ outside the $\rho$-circle.
This is explicitly shown in Fig. \ref{fig5}(b) for the longer
interval of rotating states stood out in the inset of Fig.
\ref{fig3}(a). Destabilization of the rotating kink at the borders
of such an interval gives rise to transporting states with regular
dynamics.

\begin{figure}
\begin{center}
\begin{tabular}{cc}
\includegraphics[width=3.8cm,height=3.8cm]{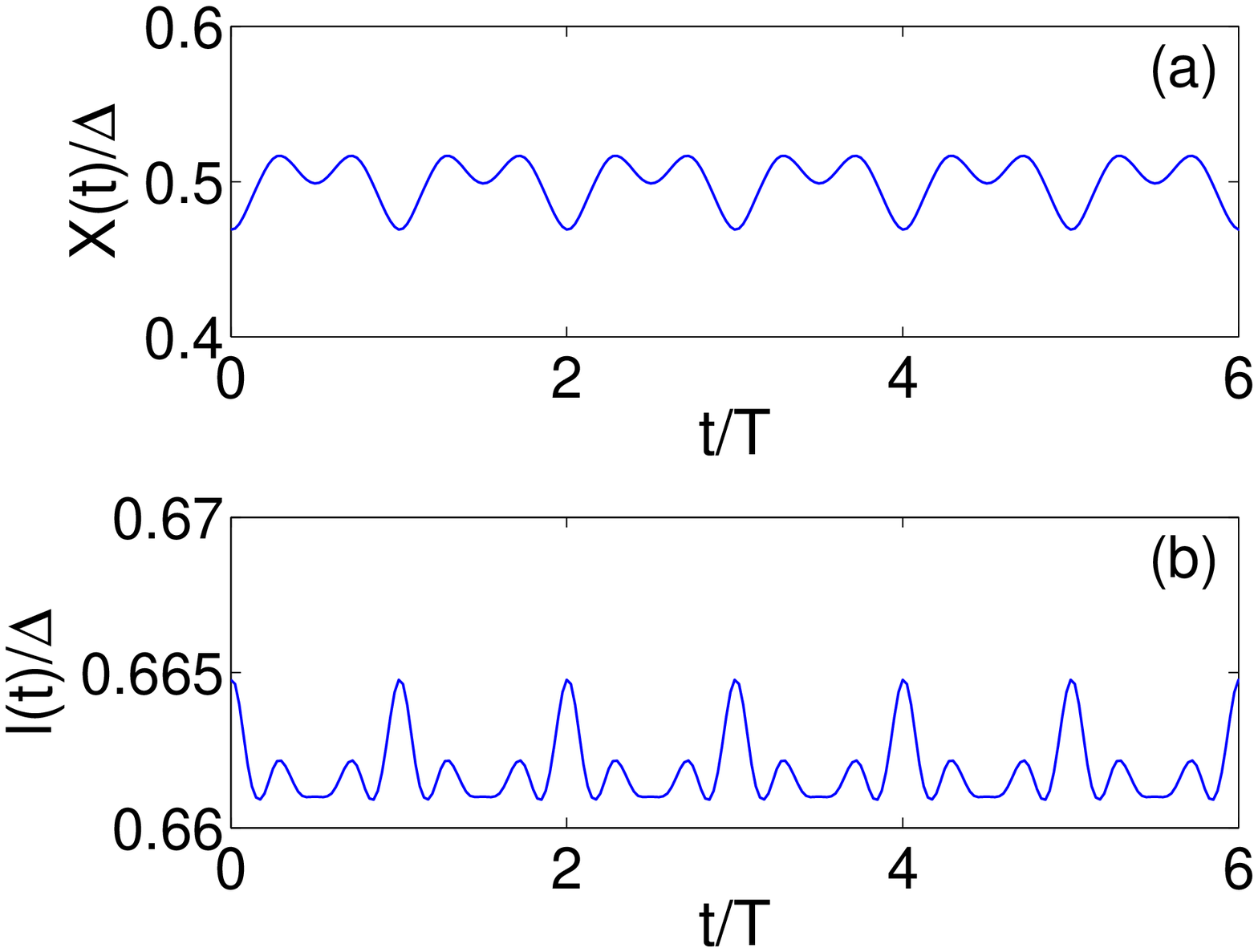} &
\includegraphics[width=3.8cm,height=3.8cm]{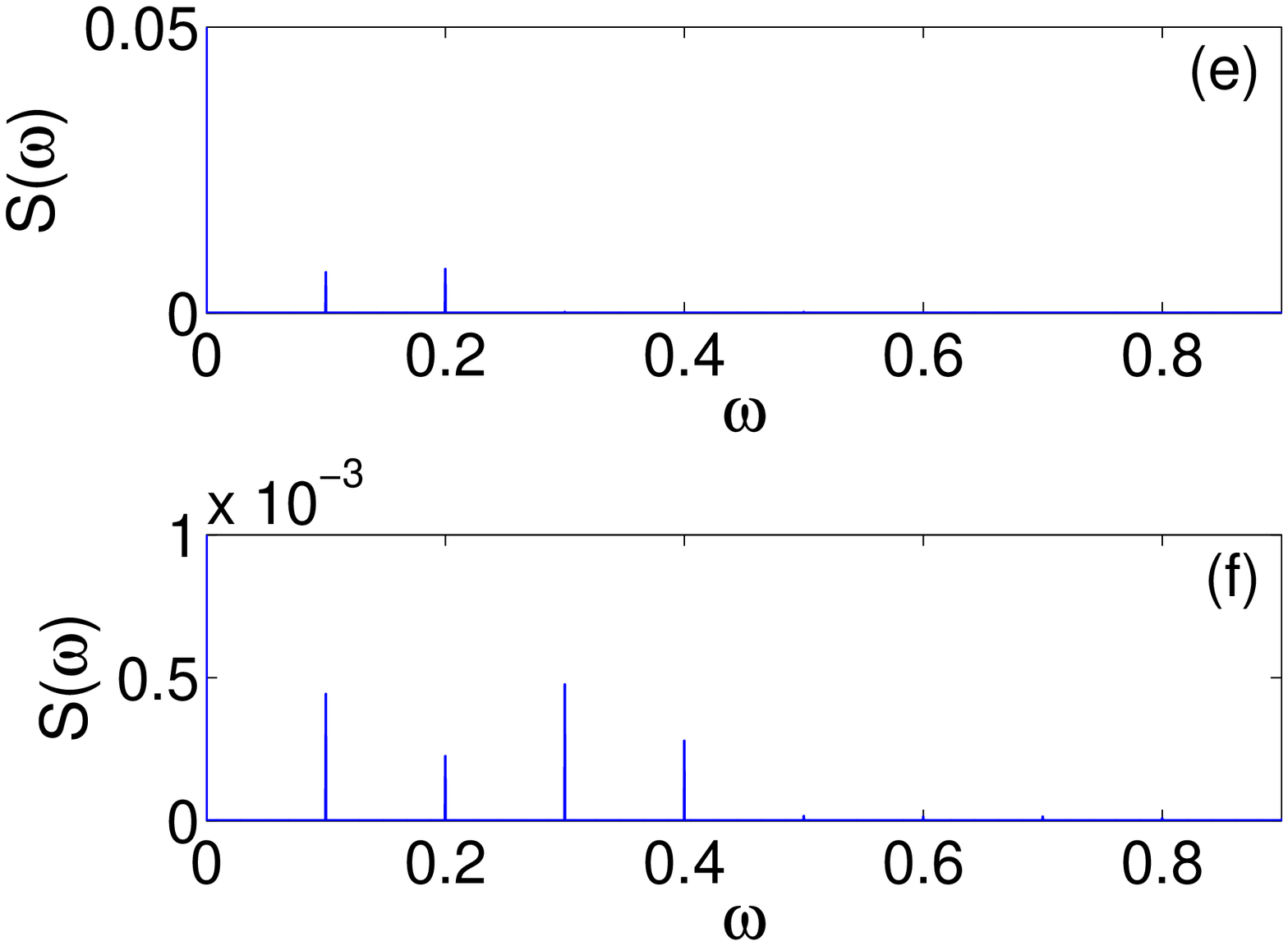}\\
\includegraphics[width=3.8cm,height=3.8cm]{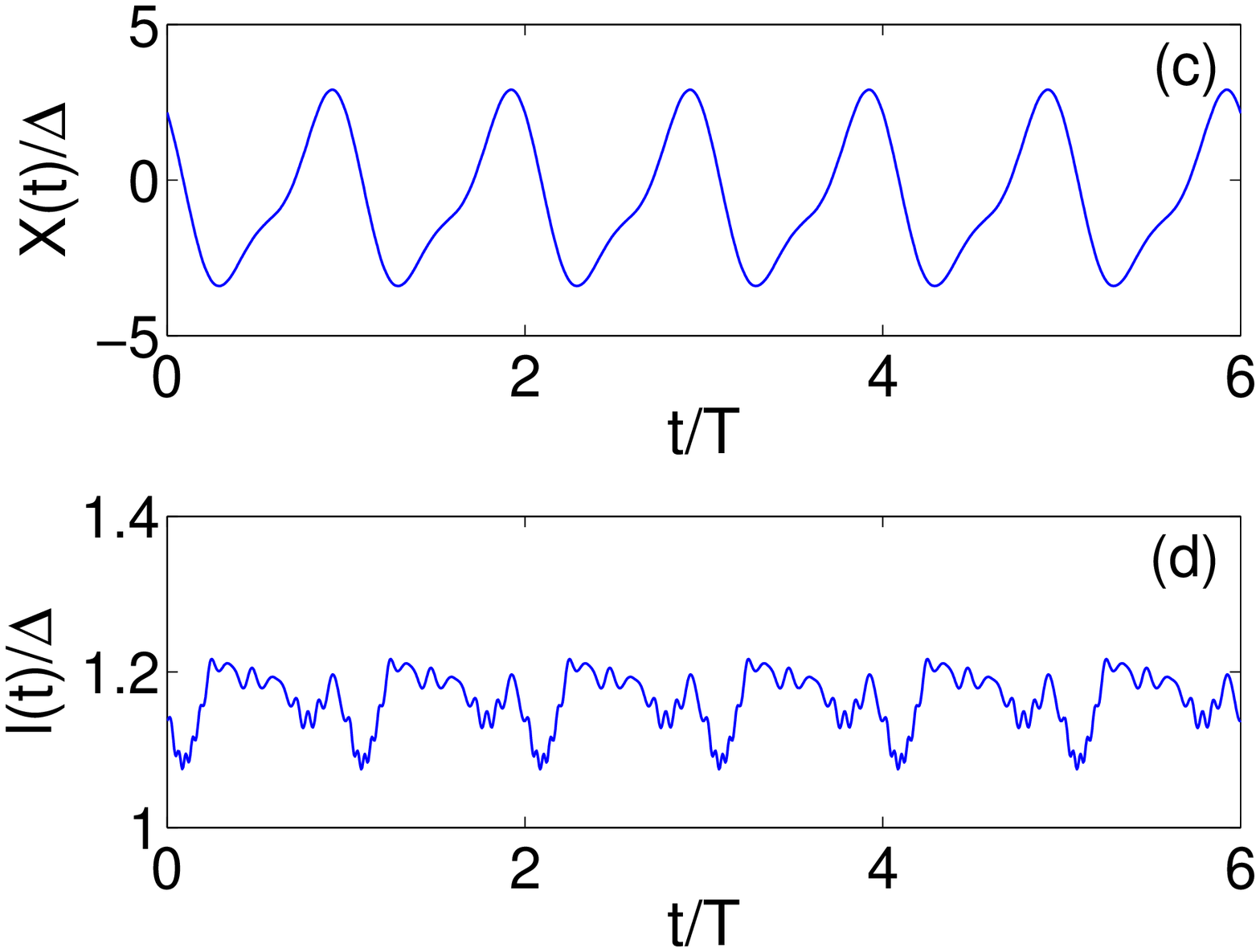} &
\includegraphics[width=3.8cm,height=3.8cm]{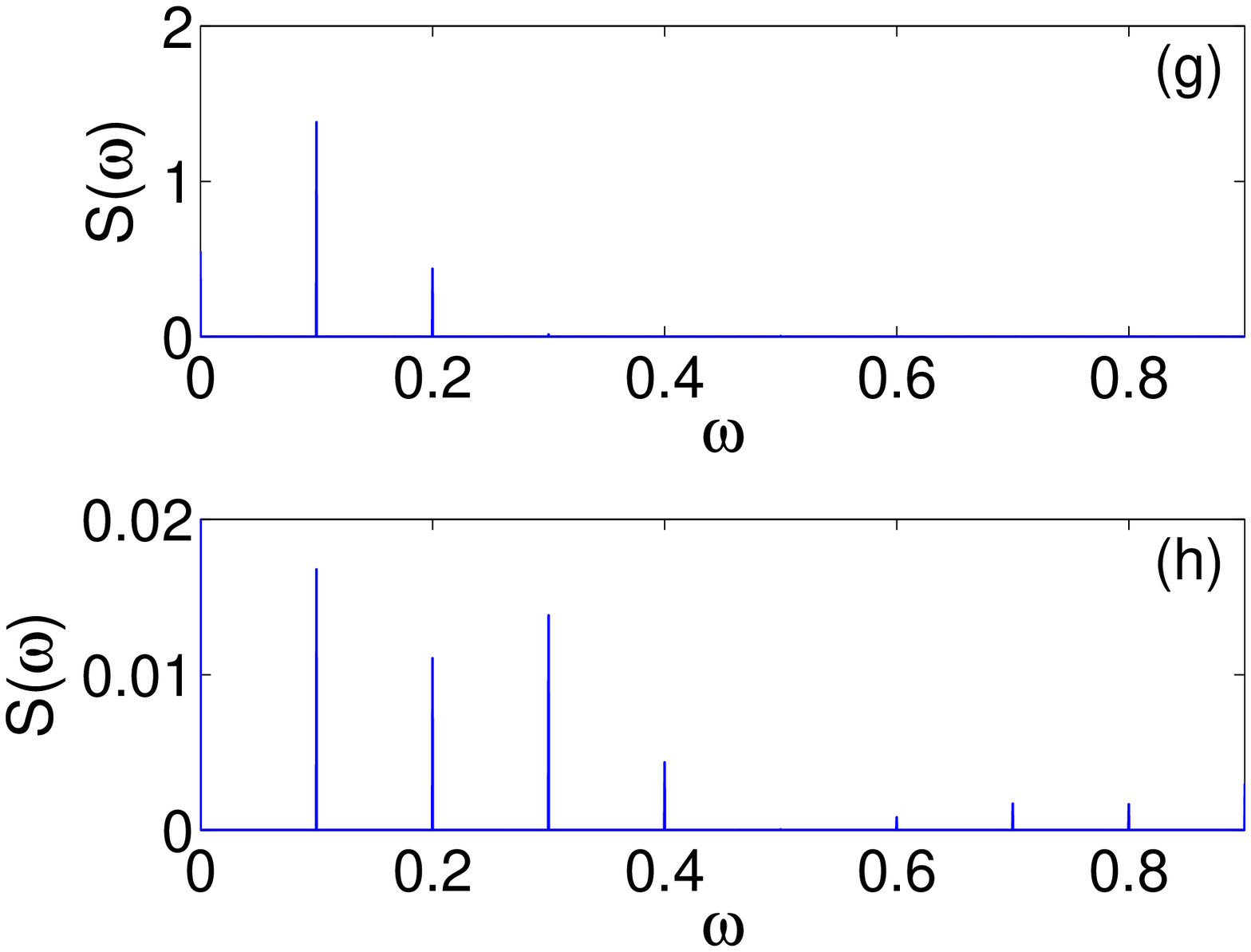}
\end{tabular}
\end{center}
\caption{Evolution of the center of mass and width of the discrete
sG kink for $\kappa=0.3$ (top panels) and $\kappa=1.2$ (bottom
panels). The former case corresponds to a pinning state whereas the
latter constitutes an example of a non-transporting rotating state.
Fourier spectra for both coordinates are depicted in the
right-hand-side panels. Numerical integration of the CC equations
(\ref{eq:Xeq})-(\ref{eq:leq}) leads to a quantitatively similar
outcome. Other parameters are $\omega=0.1, \epsilon=0.05,
\alpha=0.1$ and $\varphi=0$.} \label{fig4}
\end{figure}

\begin{figure}
\begin{center}
\begin{tabular}{c}
\includegraphics[width=6.0cm]{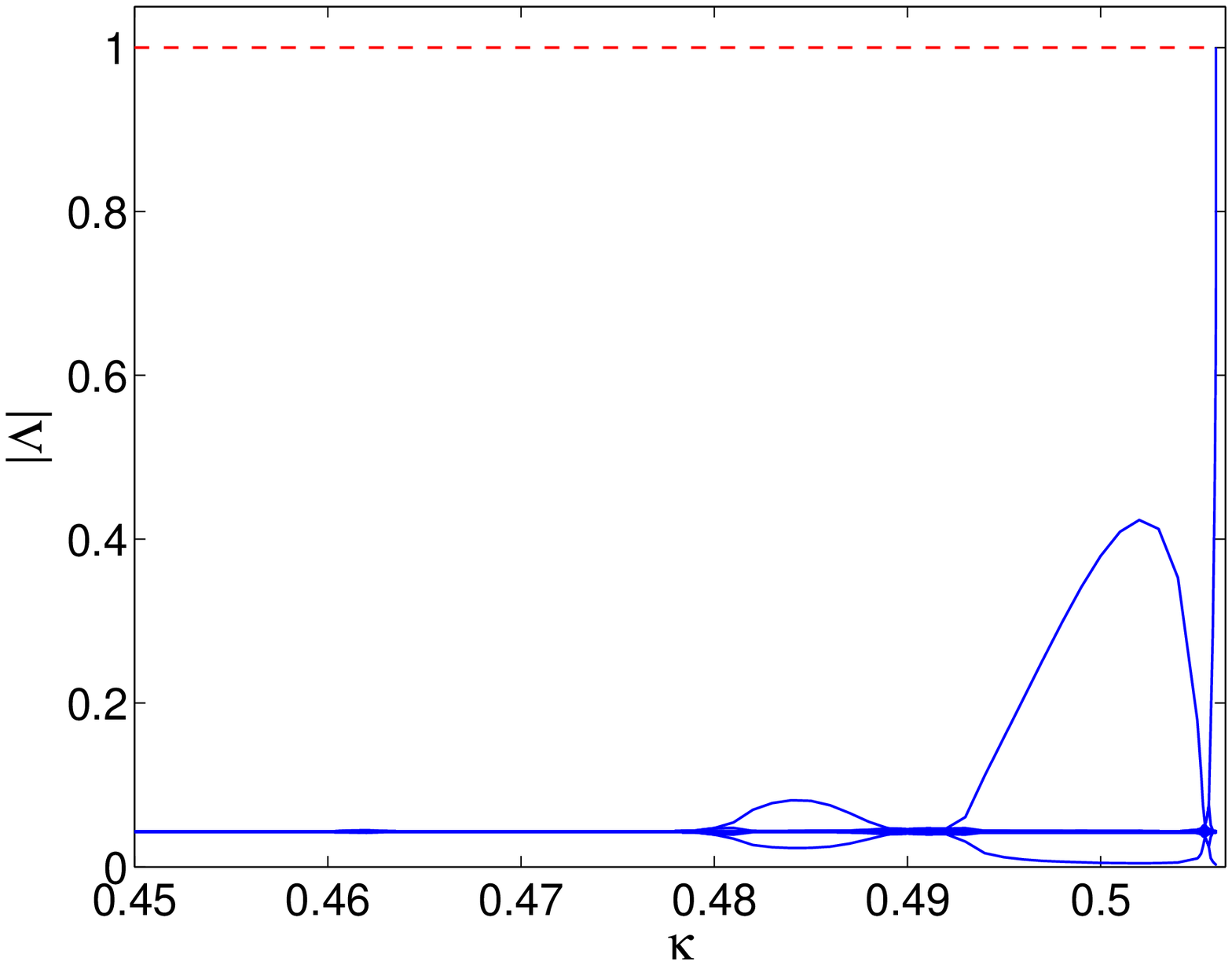} \\
\includegraphics[width=6.0cm]{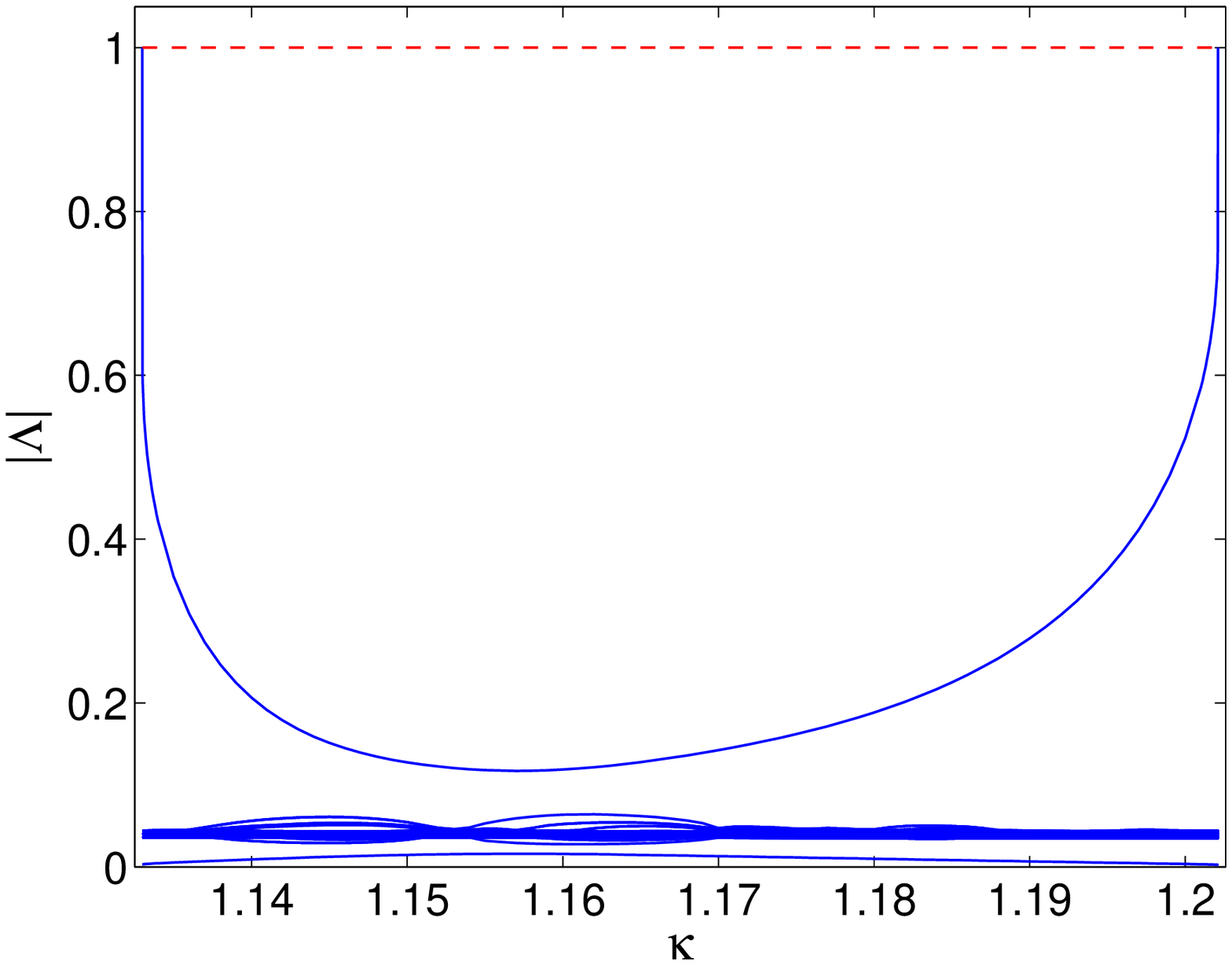}
\end{tabular}
\end{center}
\caption{Top panel: Moduli of Floquet eigenvalues versus coupling
constant for the pinning states of Fig.\ \ref{fig3}(a). Bottom
panel: Moduli of Floquet eigenvalues for the rotating states stand
out in the inset of Fig. \ref{fig3}(a).} \label{fig5}
\end{figure}

An extensive study of the existing dynamical regimes in the
parameter $(\epsilon, \kappa)$ plane is displayed in Fig.
\ref{fig6}. It is restricted to small values of $\epsilon$ because
the CC theory fails for $\epsilon>0.1$. This is in accordance with
its perturbative nature and with the fact that, in the discrete sG
system, kinks are destroyed above a critical driving by the chaotic
dynamics of the whole lattice \cite{Salerno2}. Comparison between
panels (a) and (b) of Fig. \ref{fig6} shows that the CC
approximation overestimates the green, non-transporting areas but,
in general, satisfactorily captures  the peculiar and complex
pattern in the parameter space.

\begin{figure}
\begin{center}
\includegraphics[width=6.0cm]{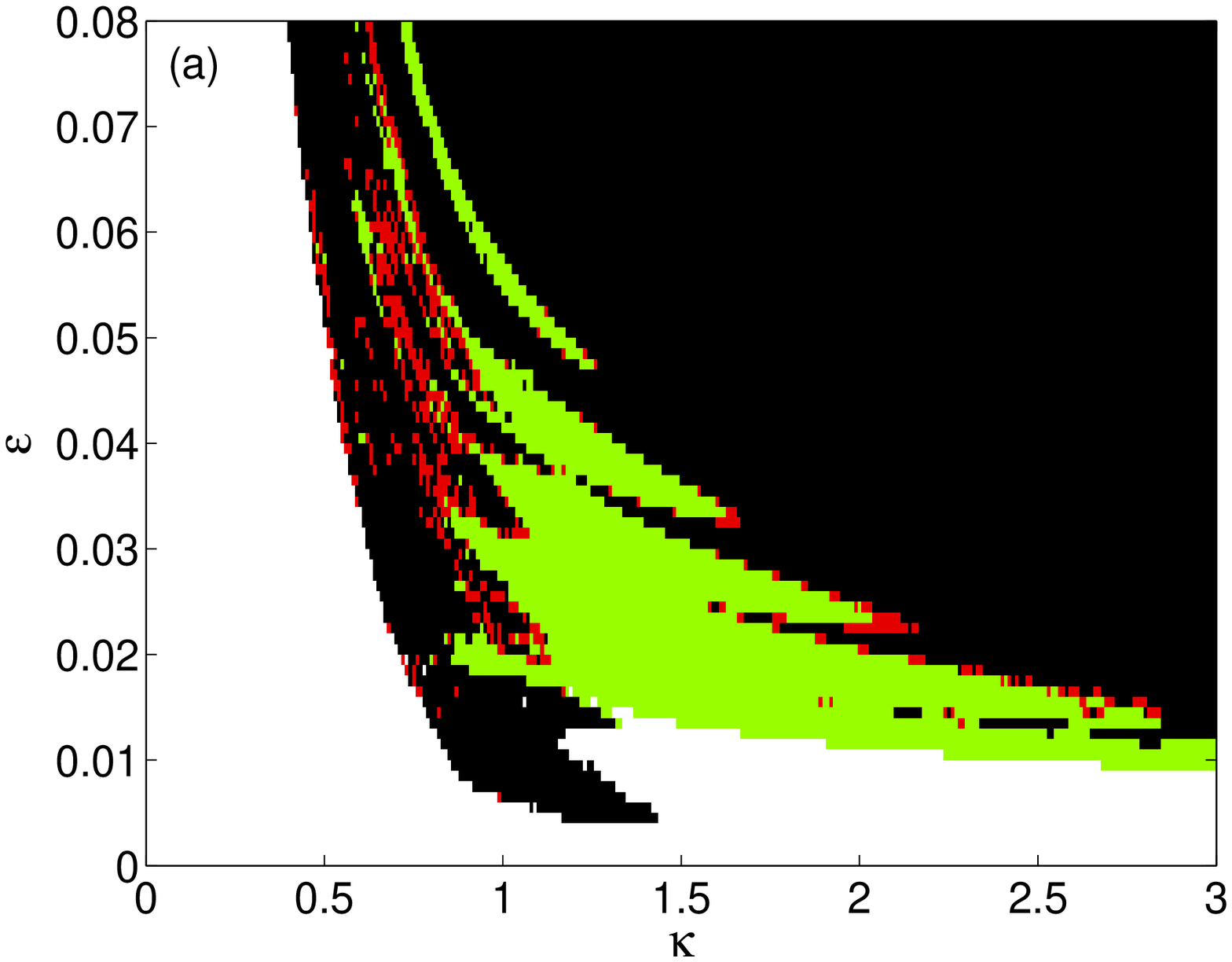}\\
\includegraphics[width=6.0cm]{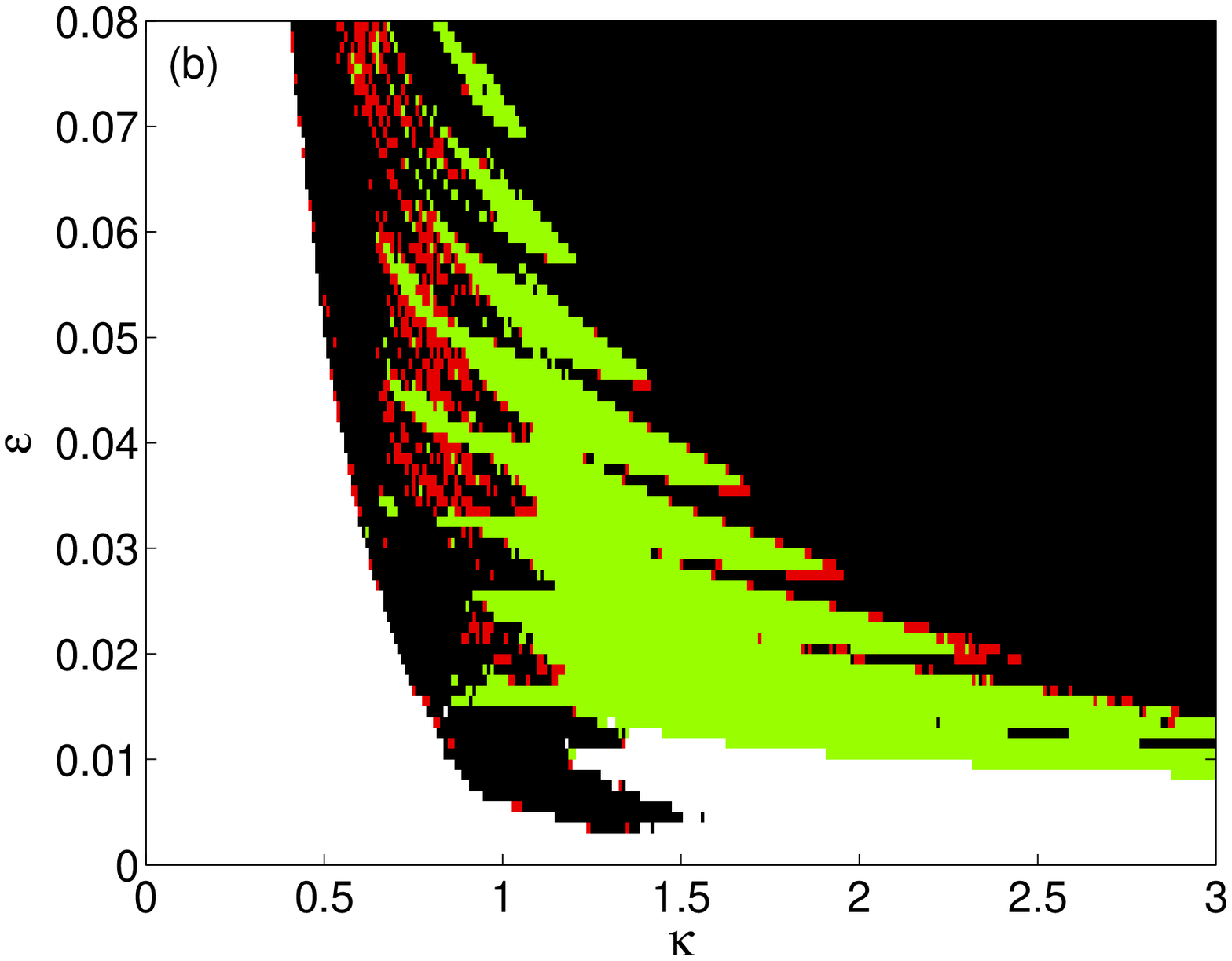} \\
\end{center}
\caption{(Color online) Dynamical regimes of the kink ratchet motion
for various values of the coupling constant, $\kappa$,  and the
driving amplitude, $\epsilon$. Panel (a) corresponds to  the results
from simulations of the discrete sG system (\ref{eq:sG}), and panel
(b) to the results from the CC equations of motion. White area: the
kink remains pinned; black: periodic or quasiperiodic regular
transport; red: chaotic diffusive transport; green: rotating states.
Other parameters are $\omega=0.1, \alpha=0.1$, $\varphi=0$. }
\label{fig6}
\end{figure}

Coming back to Fig. \ref{fig3}, another striking aspect is that
transport in the discrete system can be very effective. In fact,
although in the discrete case kinks have to overcome an energy
barrier to move through the lattice (the so-called Peierls-Nabarro
barrier), one can observe  values in Fig. \ref{fig3} of the mean
kink velocity that are an order of magnitude higher than in the
continuous case for the same parameter values. This is actually
counter-intuitive. In essence, the Peierls-Nabarro barrier decreases
exponentially with $\kappa$ \cite{Kivshar}, and therefore one could
erroneously expect the kink velocity to be a monotonous increasing
function of $\kappa$. However, resonance regimes of type
(\ref{eq:resonances}) can induce very effective transport. Such
resonant regimes have also been found in the damped
Frenkel-Kontorova model driven by a piecewise constant force
\cite{floria1996}.

Finally, we have studied the dependence of the kink velocity on the
damping. Close to the continuous limit $\langle v\rangle$ displays a
characteristic non-monotonic behaviour as  shown in Fig.
\ref{fig7}(a) for $\kappa=2$. If $\varphi=0$, in the underdamped
limit, $\langle v\rangle$ vanishes because the discrete sG system
(\ref{eq:sG}) becomes invariant under the transformation
$t\rightarrow -t$ and $\langle v\rangle$ changes its sign. On the
other hand,  large damping strongly reduces  the mobility. As a
consequence, transport is maximized for intermediate values of
damping \cite{JPA}. The CC approach (red dashed line) suitable
describes this non-monotonous behaviour and the location of the
peak.

As the coupling is decreased,  kinks lose  their mobility quickly
and the ratchet effect persists only for small values of damping.
For instance, in Fig. \ref{fig7}(b), one can see that for $\kappa=1$
net transport exists basically  for $\alpha \lesssim 0.1$. In that
region, $\langle v\rangle$ no longer varies smoothly on $\alpha$ but
instead shows the characteristic staircase structure of discrete
systems. Although the CC approximation gives  the mobility range and
the order of magnitude of $\langle v\rangle$ correctly it fails, for
small values of damping, in the prediction of a current reversal,
which is not observed in the sG system (\ref{eq:sG}).

\begin{figure}
\begin{center}
\includegraphics[width=6.0cm]{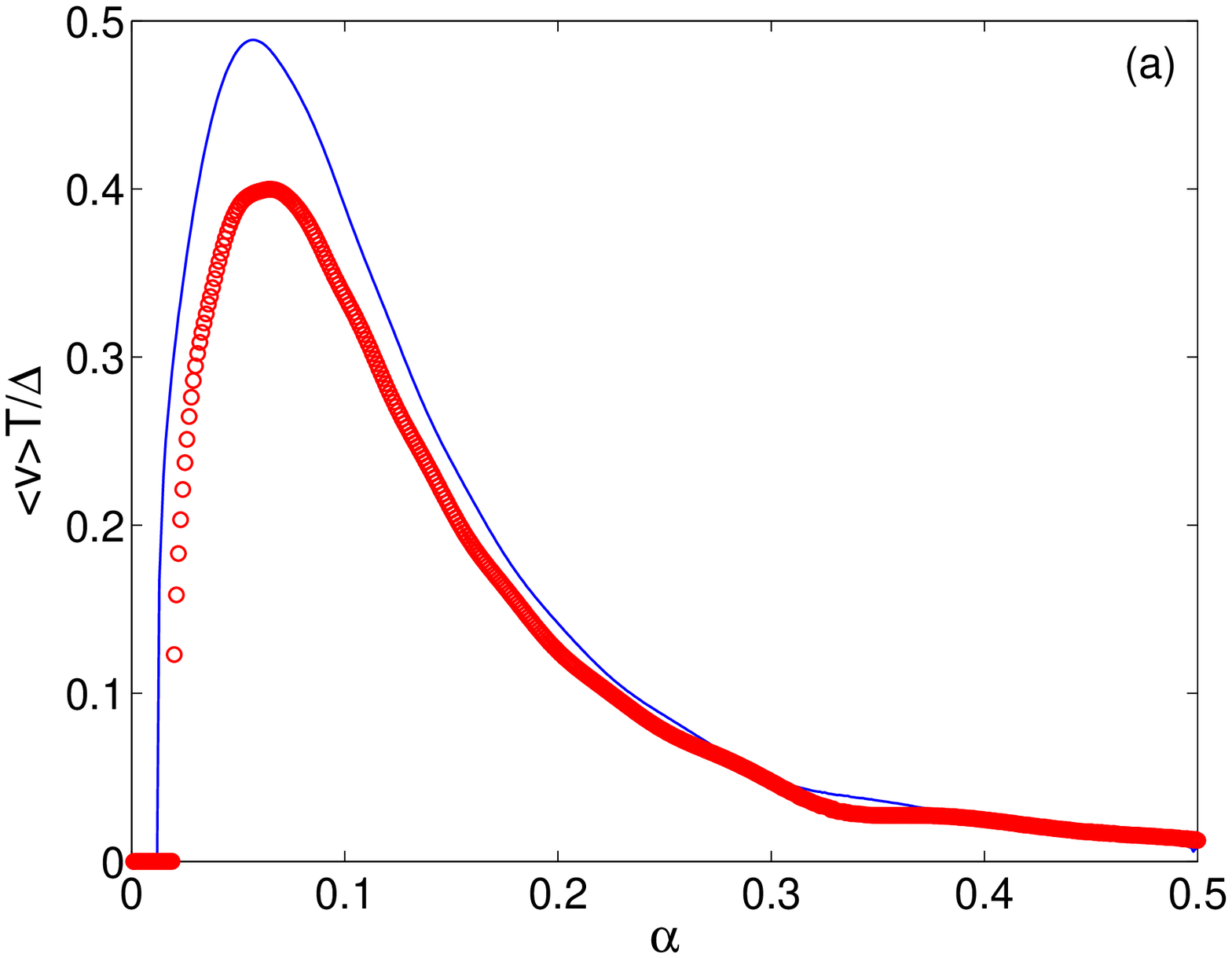}\\
\includegraphics[width=6.0cm]{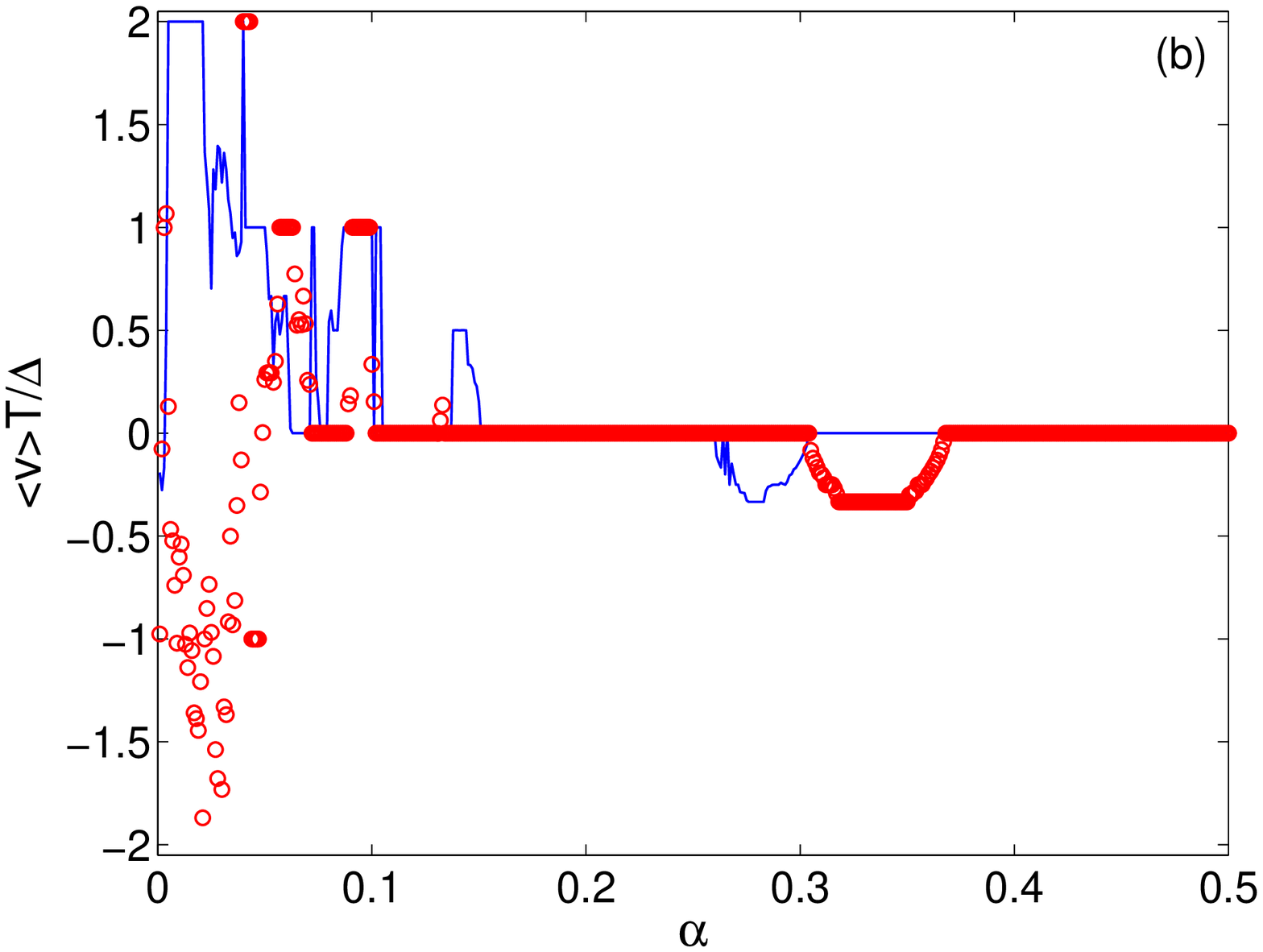} \\
\end{center}
\caption{(Color online) Dependence of the  normalized average kink
velocity on the damping for $\kappa=2$ (a), and $\kappa=1$ (b). The
blue solid line represents results from the simulations of the
discrete sG equation (\ref{eq:sG}) while the red circles correspond
to numerical integration of the CC equations. Other parameters:
$\omega=0.1$, $\epsilon=0.05$ and $\varphi=0$.} \label{fig7}
\end{figure}

\section{Conclusions}
We have developed a CC theory for kink ratchets in the damped
discrete sine-Gordon equation driven by a bi-harmonic force.
Inspired in the Rice {\it ansatz} used in the continuum sine-Gordon
equation \cite{Rice},   a discrete {\it ansatz} is suggested with
two collective coordinates, the center and the width of the soliton,
as an approximated solution of our working discrete non-linear
equation. The evolution of these two collective coordinates has been
obtained by means of  the  Generalized Travelling Wave Method. The
resulting CC theory  explains the mechanism underlying the discrete
soliton ratchet (the bi-harmonic force with zero average acts on the
discrete sG equation, whereas an effective force with, in general,
non-zero average acts on the center of kink and causes its motion)
and captures the distinctive features of kink motion in discrete
systems: namely the existence of a depinning threshold; the
piece-wise dependence of the mean velocity on system parameters; and
the complicated structure of subspaces of transporting and
non-transporting regimes in parameter space.

The numerical study shows that the theory agrees well with the
results obtained from simulations of the discrete sG equation close
to the continuous limit ($\kappa > 1$) and for small amplitudes of
the bi-harmonic force ($\epsilon<0.1$). This is consistent with the
perturbative nature of the CC approach. When $\kappa\le 1$, the
agreement is not good from a quantitative point of view but the CC
approximation still  provides a reasonable qualitative description
of what happens in the very discrete limit.

Particular attention has been devoted to the investigation of the
intriguing shape of areas  corresponding to different dynamical
regimes in phase space. Comparison with exact numerical results of
the discrete sG system reveals that the CC approach satisfactorily
captures the peculiar and complex structure of subspaces in the
parameter plane $(\epsilon, \kappa)$.

Further development of the theory is needed in order to extend the
method to the case of kink ratchets in asymmetric double sG
potentials. In fact, it is easy to verify that the system of CC
equations (\ref{eq:Xeq})-(\ref{eq:leq}) loses its dependence on the
asymmetry parameter, $\lambda$, when taking the continuous limit.
The challenge is to find an adequate \textit{ansatz} for the
asymmetric discrete sG system that leads to a tractable system of
ODEs for the CCs.

\appendix

\section{Approximations for the sums}
All the sums $I_j\; (j=1,...,13)$ defined in the set of equations
(\ref{eq:I1})-(\ref{eq:I13}) are periodic functions of $X(t)$.
Therefore, the sums can be expressed as Fourier series and
approximated by the first term of each series:

\begin{widetext}
\be %
I_1 = 16 \pi^3 \kappa^{3/2} l^3 \sum_{n=1}^{\infty}
\frac{n^2}{\sinh(n\pi^2 \sqrt{\kappa}\, l)} \sin(2 n\pi
\sqrt{\kappa} X) \approx \frac{16 \pi^3 \kappa^{3/2}
l^3}{\sinh(\pi^2 \sqrt{\kappa}\, l)} \sin(2\pi \sqrt{\kappa}\, X),
\ee %
\be%
 I_2=-4 \sqrt{\kappa}\, l-16 \pi^2 \kappa\, l^2
\sum_{n=1}^{\infty} \frac{n[1-\frac{n\pi^2 \sqrt{\kappa}\,l}{2}
\coth(n\pi^2 \sqrt{\kappa}\,l)]}{\sinh(n\pi^2\sqrt{\kappa}\,l)}
\sin(2 n\pi \sqrt{\kappa} X)\approx -4 \sqrt{\kappa}\, l,
\ee %
\bea %
I_3&=& \sum_{n=1}^{\infty} \frac{ -4 \pi \sqrt{\kappa}\,
l}{\sinh( n\pi^2\sqrt{\kappa}\, l)} \Bigg[ 2+n^2\pi^4\kappa\, l^2+ 2
n \pi^2 \sqrt{\kappa}\, l\left( -2 \coth(n\pi^2 \sqrt{\kappa}\, l)
+\frac{ n\pi^2\sqrt{\kappa}\, l}{ \sinh^2( n\pi^2\sqrt{\kappa}\, l)}
\right) \Bigg]  \sin(2 n\pi \sqrt{\kappa} X)
\nonumber \\
&\approx& -\frac{4 \pi \kappa^{1/2} l}{\sinh(\pi^2 \sqrt{\kappa}\,
l)} \left[ 2+\pi^4 \kappa l^2 -4 \pi^2 \sqrt{\kappa}\, l \coth
(\pi^2 \sqrt{\kappa}\, l) +\frac{2 \pi^4 \kappa l^2}{\sinh^{2}
(\pi^2 \sqrt{\kappa}\, l)}\right] \sin(2 \pi \sqrt{\kappa}\,X),
\end{eqnarray}
\be %
I_4=8 \sqrt{\kappa}\, l+ 16\pi^2 \kappa l^2 \sum_{n=1}^{\infty}
\frac{n}{\sinh(n\pi^2 \sqrt{\kappa}\, l)} \cos(2 n\pi \sqrt{\kappa}
X)
\approx 8  \sqrt{\kappa}\, l,
\ee %
\begin{eqnarray}
I_5&=& 8\pi \sqrt{\kappa}\, l \sum_{n=1}^{\infty}
\frac{1-n\pi^2\sqrt{\kappa}\, l \coth(n\pi^2\sqrt{\kappa}\,
l)}{\sinh(n\pi^2\sqrt{\kappa}\, l)} \sin(2 n\pi \sqrt{\kappa} X)
\nonumber \\ &\approx& \frac{8 \pi \kappa^{1/2} l
[1-\pi^2\sqrt{\kappa}\, l \coth (\pi^2 \sqrt{\kappa}\, l) ]}
{\sinh(\pi^2 \sqrt{\kappa}\, l)} \sin(2 \pi \sqrt{\kappa}\, X),
\eea %
\be %
I_6=2 \pi \sqrt{\kappa}\, l+4 \pi \sqrt{\kappa}\, l
\sum_{n=1}^{\infty} \frac{\cos(2 n\pi \sqrt{\kappa} X)}{\cosh(n\pi^2
\sqrt{\kappa}\, l)}\approx 2 \pi \sqrt{\kappa}\, l,
\ee %
\bea %
I_7&=& I_6-2 \pi \sqrt{\kappa}\, l-\frac{4}{3} \pi \sqrt{\kappa}\, l
\sum_{n=1}^{\infty} \frac{3+8n^2 \pi^2\kappa\,l^2-16 n^4 \pi^4
\kappa^2\, l^4}{\cosh(n \pi^2 \sqrt{\kappa}\, l)} \cos(2 n\pi
\sqrt{\kappa} X) \nonumber \\ &\approx& \frac{32 \pi^3 \kappa^{3/2}
l^3(2 \pi^2 \kappa l^2-1)} {3 \cosh(\pi^2 \sqrt{\kappa}\, l)} \cos(2
\pi\sqrt{\kappa}\, X),
\eea %
\be %
I_8= I_1+32 \sum_{n=1}^{\infty}  \frac{n^2\pi^3 \kappa^{3/2}\, l^3
(1+n^2\pi^2\kappa \, l^2)}{\sinh(n\pi^2 \sqrt{\kappa}\, l)}
\sin(2\pi \sqrt{\kappa}\,X)) \approx \frac{16\pi^3 \kappa^{3/2} l^3
(3+2 \pi^2 \kappa l^2)}{\sinh(\pi^2 \sqrt{\kappa}\, l)} \sin(2\pi
\sqrt{\kappa}\,X) ,
\ee %
\begin{eqnarray}
I_{9}&=&  -\pi^2 \sqrt{\kappa}\, l-\sum_{n=1}^{\infty} \frac{\pi^2
\sqrt{\kappa}\, l}{2 \sinh^4(n\pi^2 \sqrt{\kappa}\, l)} \Bigg[
(-6+23 n^2 \pi^4 \kappa\, l^2) \cosh(n\pi^2\sqrt{\kappa}\, l)+(6+n^2
\pi^4\kappa\, l^2) \cosh(3n\pi^2\sqrt{\kappa}\, l) \nonumber \\ &&
-6 n \pi^2 \sqrt{\kappa}\, l \Big( 5\sinh(n\pi^2 \sqrt{\kappa}\, l)+
\sinh(3 n\pi^2 \sqrt{\kappa}\, l) \Big)\Bigg] \cos(2 n\pi
\sqrt{\kappa} X) \approx  -\pi^2 \sqrt{\kappa}\, l,
\eea %
\bea %
I_{10}&=& \frac{2}{3} \pi^2 \sqrt{\kappa}\, l-\sum_{n=1}^{\infty}
\frac{\pi^2 \sqrt{\kappa}\, l}{\sinh^3(n\pi^2 \sqrt{\kappa}\, l)}
\Bigg[2 n \pi^2 \sqrt{\kappa}\, l \Big( 3+ \cosh(2
n\pi^2\sqrt{\kappa}\, l)\Big) -4 \sinh(2n\pi^2\sqrt{\kappa}\,
l)\Bigg]  \cos(2 n\pi \sqrt{\kappa} X) \nonumber \\
&\approx& \frac{2}{3} \pi^2 \sqrt{\kappa}\, l,
\end{eqnarray}
\be I_{11}= -2 \pi^2 \sqrt{\kappa}\, l \sum_{n=1}^{\infty}
\frac{\tanh(n\pi^2\sqrt{\kappa}\, l)} {\cosh(n\pi^2\sqrt{\kappa}\,
l)} \sin(2 \pi\sqrt{\kappa}\, X) \approx -\frac{2 \pi^2 \kappa^{1/2}
l
 \tanh(\pi^2 \sqrt{\kappa}\, l)}{\cosh(\pi^2 \sqrt{\kappa}\, l)} \sin(2 \pi\sqrt{\kappa}\, X),
\ee
\begin{eqnarray}
I_{12}&=& I_{11}+ \sum_{n=1}^{\infty}  \frac{2\pi \sqrt{\kappa}\,
l}{3\cosh(n\pi^2 \sqrt{\kappa}\, l)} 
\Bigg[(3 \pi +8 n^2 \pi^3 \kappa\, l^2-16 n^4 \pi^5\kappa^2 \, l^4)
\tanh( n\pi^2\sqrt{\kappa}\, l)+16 n\pi \sqrt{\kappa}\, l(4 n^2
\pi^2\kappa\, l^2-1) \Bigg]   \nonumber \\&&\times \sin(2 n\pi \sqrt{\kappa} X) 
\approx \frac{16 \pi^2 \kappa l^2} {3\cosh(\pi^2 \sqrt{\kappa}\, l)}
\Big[ -2+8 \pi^2 \kappa l^2 + \pi^2
\sqrt{\kappa}\, l (1-2\pi^2\kappa\, l)\tanh (\pi^2 \sqrt{\kappa}\, l)\Big] \sin(2 \pi \sqrt{\kappa}\,X), \\
I_{13}&=& I_2-8 \sqrt{\kappa}\, l-\sum_{n=1}^{\infty}  \frac{16 n
\pi^2 \kappa\, l^2}{\sinh(n\pi^2 \sqrt{\kappa}\, l)} \Bigg[2 (1 +2
n^2 \pi^2 \kappa\, l^2)- n \pi^2 \sqrt{\kappa}\, l
(1+n^2\pi^2\kappa\, l^2) \coth( n\pi^2\sqrt{\kappa}\, l) \Bigg]
\cos(2 n\pi \sqrt{\kappa} X) \nonumber \\ &\approx& -12
\sqrt{\kappa}\, l.
\end{eqnarray}
\end{widetext}

Clearly, since $l(t)$ is an oscillating function around $l_0\approx
1$, the sums $I_2,I_4,I_6,I_{9},I_{10}$ and $I_{13}$ are
$\mathcal{O}(\kappa^{1/2})$. In Fig. \ref{fig8},  we have plotted
the amplitudes of the remaining series, $I_8$, $I_7$, $I_3$, $I_1$,
$I_5$, $I_{11}$ and $I_{12}$ from top to bottom, versus the coupling
parameter $\kappa$ in order to weigh up the importance of each term
in Eqs.(\ref{eq:Xeq}) and (\ref{eq:leq}). These amplitudes have been
evaluated at $l(t)\approx 1$ due to the small size of the
fluctuations of the kink width. From Fig. \ref{fig7}, it is clear
that $I_5$, $I_{11}$ and $I_{12}$ are much smaller than the rest and
can be dropped in Eqs. (\ref{eq:Xeq})-(\ref{eq:leq}).

\begin{figure}
\begin{center}
\includegraphics[width=8.0cm]{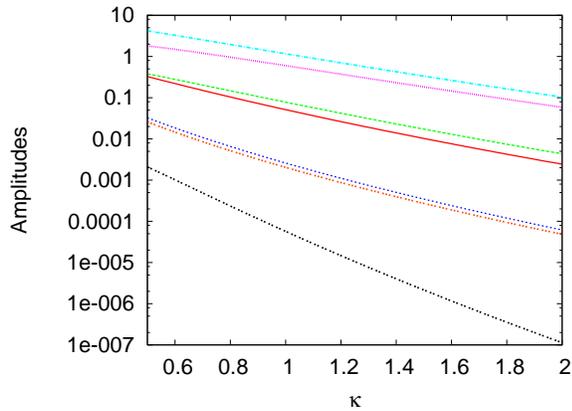}
\end{center}
\caption{(Color online) Amplitudes of the series $I_8$, $I_7$,
$I_3$, $I_1$, $I_5$, $I_{11}$ and $I_{12}$ (from top to bottom)
versus coupling. The amplitudes have been evaluated at $l(t)\approx
1$ due to the small size of the fluctuations of $l(t)$.}
\label{fig8}
\end{figure}

\begin{acknowledgements}
N.R.Q. acknowledges financial support through grant FIS2011-24540
from the Ministerio de Econom\'{\i}a y Competitividad (Spain),
grants FQM207, P09-FQM-4643 from Junta de Andaluc\'{\i}a (Spain),
and a grant from the Humboldt Foundation through a Research
Fellowship for Experienced Researchers SPA 1146358 STP.
\end{acknowledgements}


\begin{thebibliography}{88}

\bibitem{Marchesoni} F.\ Marchesoni, Phys.\ Rev.\ Lett.\ {\bf 77}, 2364
(1996).

\bibitem{Niurka1} M. Salerno and N. R.\ Quintero,
Phys. Rev. E, {\bf 65} 025602(R) (2002).

\bibitem{Costantini} G.\ Costantini, F.\ Marchesoni and
M.\ Borromeo, Phys.\ Rev.\ E {\bf 65}, 051103 (2002).


\bibitem{Ratchet} P.\ Reimann  Phys.\ Rep.\  {\bf 361}, 57
(2002); P. H\"anggi and R. Bartussek, in {\it Lecture notes in
Physics}, Ed.s J. Parisi et al., Springer, Berlin, {\bf 476} (1996).


\bibitem{Salerno1}  M. Salerno and Y. Zolotaryuk.
Phys.\ Rev.\ E {\bf 65}, 056603 (2002).

\bibitem{Flach}  S.\ Flach, Y.\ Zolotaryuk, A.\ E.\ Miroshnichenko,
and M.\ V.\ Fistul. Phys.\ Rev.\ Lett.\ {\bf 88}, 184101 (2002).

\bibitem{Carapella} G. Carapella and G. Costabile,
Phys. Rev. Lett. {\bf 87}, 077002 (2001); G. Carapella,\ Phys.\ Rev.
\ B {\bf 63}, 054515 (2001).

\bibitem{Beck} M. Beck, E. Goldobin, M. Neuhaus, M. Siegel, R.
Kleiner and D. Koelle, Phys. Rev. Lett. {\bf 95}, 090603 (2005).

\bibitem{Trias} E. Tr\'{\i}as, J.J. Mazo, F. Falo and T.P. Orlando, Phys.
Rev. E {\bf 61}, 2257 (2000).


\bibitem{Shalom} D.E. Shalom and H. Pastoriza, Phys. Rev. Lett. {\bf 94}, 177001
(2005).

\bibitem{Ustinov04} A.\ V.\ Ustinov, C. Coqui, A. Kemp, Y. Zolotaryuk and M. Salerno, Phys.\ Rev.\ Lett.\ {\bf 93},
087001 (2004).

\bibitem{Niurka2} L.\ Morales-Molina,N.\ R.\ Quintero, F.\ G.\
Mertens and A.\ S\'anchez, Phys.\ Rev.\ Lett.\ {\bf 91}, 234102
(2003).

\bibitem{Niurka3} N.\ R.\ Quintero, B. S\'anchez-Rey and M.\ Salerno, Phys. Rev. E
{\bf 72}, 016610 (2005); N.\ R.\ Quintero, R.\ Alvarez-Nordase and
F.\ G.\ Mertens, Phys.\ Rev.\ E\ {\bf 80}, 016605 (2009).

\bibitem{moralesmolina2004} L. Morales-Molina, F. G. Mertens and A. S\'{a}nchez,
 Eur.\ Phys. J. B {\textbf 37}, 79 (2004).

\bibitem{moralesmolina2005} L. Morales-Molina, F. G. Mertens and A. S\'{a}nchez,
 Phys.\ Rev. E {\textbf 72}, 016612 (2005).

\bibitem{Salerno2}  Y. Zolotaryuk and M. Salerno
Phys.\ Rev.\ E {\bf 73}, 066621 (2006).

\bibitem{Cuevas} J. Cuevas, B. S\'anchez-Rey and M. Salerno, Phys.
Rev. E {\bf 82}, 016604 (2010).

\bibitem{Gorbach} A. Gorbach, S. Denisov and S. Flach, Opt. Lett. {\bf 31}, 1702
(2006).

\bibitem{Chacon} P.J. Mart\'{\i}nez and R. Chacon, Phys. Rev. Lett. {\bf 100}, 144101
(2008).

\bibitem{yang} Y. Yang, et. al., Eur. Phys. Lett. {\bf 93}, 16001
(2001).


\bibitem{Zolotaryuk}  Y. Zolotaryuk, Phys.\ Rev.\ E {\bf 86}, 026604 (2012).

\bibitem{Braun1998} Oleg M.\ Braun and Yuri S. Kivshar,
Phys.\ Rep.\ {\bf 306}, 2 (1998).

\bibitem{Kivshar} Oleg M.\ Braun and Yuri S. Kivshar, {\it The Frenkel-Kontorova Model:
Concepts, Methods and Applications}, Springer-Verlag,
Berlin-Heidelberg, 2004.

\bibitem{sGbook} J. Cuevas-Maraver, P.G. Kevrekidis, and F. Williams, eds. {\it The sine-Gordon Model and its Applications.
From Pendula and Josephson Junctions to Gravity and High Energy
Physics}, Springer-Verlag, Berlin-Heidelberg. In press.

\bibitem{Falo} F. Falo, P.J. Mart\'{\i}nez, J.J. Mazo and S. Cilla,
Europhys. Lett. {\bf 45}, 700 (1999)

\bibitem{usti1993} A. V. Ustinov, M. Cirillo and B. Malomed,
 Phys.\ Rev.\ B {\bf 47}, 8357 (1993).

\bibitem{ms85} M. Salerno, Physica {\bf D17}, 227 (1985).

\bibitem{Rice} M.J. Rice and E.J. Mele, Solid State Commun. {\bf
35}, 487 (1980); M. Salerno and A.C. Scott Phys. Rev. B {\bf 26},
2474 (1982).

\bibitem{cisneros2008} L. A. Cisneros and A. A. Minzoni,
Physica D {\bf 237}, 50 (2008).

\bibitem{Ustinov92}  A.\ V.\ Ustinov, T. Doderer, R.P. Huebener, N.F. Pedersen, B. Mayer  and
U.A. Oboznov, Phys.\ Rev.\ Lett.\ {\bf 69}, 1815 (1992).

\bibitem{Mertens} F.G. Mertens, H.J. Schnitzer and A. R. Bishop,
Phys. Rev. B, {\bf 56}, 2510 (1997).

\bibitem{Kamppeter} T. Kamppeter et al.,
Phys. Rev. B, {\bf 59}, 11349 (1999); T. Kamppeter et al., Eur.
Phys. J. B, {\bf 7}, 607 (1999).

\bibitem{Niurka4} N.R. Quintero, A. S\'anchez and F.G. Mertens, Phys. Rev. E, {\bf 62}, 5695
(2000).

\bibitem{quintero2013} N. R. Quintero, J.A. Cuesta and R. \'Alvarez-Nodarse, Phys. Rev. E, {\bf 81}, 030102 (2010).

\bibitem{cuesta2013} J.A. Cuesta, N. R. Quintero and R. \'Alvarez-Nodarse, Phys. Rev. X \textbf{3}, 041014 (2013).

\bibitem{Marin} J.L. Mar\'{\i}n, F. Falo, P.J. Mart\'{\i}nez and L.M. Flor\'{\i}a, Phys. Rev. E {\bf 63}, 066603 (2001).

\bibitem{floria1996} L.M. Flor\'{\i}a and J. J. Mazo, Adv. Phys. {\bf 45}, 505  (1996).

\bibitem{JPA} N.R. Quintero, J.A. Cuesta and R. \'Alvarez-Nodarse, J. Phys. A: Math. Theor. \textbf{44}, 425205 (2011).

\end{thebibliography}
\end{document}